\documentclass[aps,prx,twocolumn,notitlepage,superscriptaddress]{revtex4-2}
\usepackage{bm} \usepackage{graphicx} \usepackage{amsmath,amsthm,stmaryrd,amssymb} 
\usepackage{color}
\usepackage{hyperref}

\DeclareMathOperator*{\Tr}{Tr}
\DeclareMathOperator*{\ave}{\mathbb E}
\DeclareMathOperator*{\argmin}{argmin}
\DeclareMathOperator*{\argmax}{argmax}
\newcommand{\ket}[1]{\vert{#1}\rangle}
\newcommand{\bra}[1]{\langle{#1}\vert}

\newcommand{\bs}[1]{\boldsymbol{#1}}

\newtheorem{lemma}{Lemma}
\newtheorem{thm}{Theorem}

\begin{document}

\title{Generalization in Quantum Machine Learning: \\
a Quantum Information Perspective}

\author{Leonardo Banchi} \email{leonardo.banchi@unifi.it}
\affiliation{Department of Physics and Astronomy, University of Florence,
via G. Sansone 1, I-50019 Sesto Fiorentino (FI), Italy}
\affiliation{ INFN Sezione di Firenze, via G. Sansone 1, I-50019, Sesto Fiorentino (FI), Italy }

\author{Jason Pereira}
\affiliation{Department of Computer Science, University of York, York YO10 5GH, UK}
\affiliation{Department of Physics and Astronomy, University of Florence,
via G. Sansone 1, I-50019 Sesto Fiorentino (FI), Italy}

\author{Stefano Pirandola}
\affiliation{Department of Computer Science, University of York, York YO10 5GH, UK}

\begin{abstract}
	Quantum classification and hypothesis testing are two tightly related subjects, the main difference being that the former is data driven: how to assign to quantum states $\rho(x)$ the corresponding class $c$ (or hypothesis) is learnt from examples during training, where $x$ can be either tunable experimental parameters or classical data ``embedded'' into quantum states.  Does the model generalize? This is the main question in any data-driven strategy, namely the ability to predict the correct class even of previously unseen states. 
	Here we establish a link between quantum machine learning classification and quantum hypothesis testing (state and channel discrimination) and then show that the accuracy and generalization capability of quantum 
	classifiers depend on the (R\'enyi) mutual informations $I(C{:}Q)$ and $I_2(X{:}Q)$
	between the quantum state space $Q$ and the classical parameter space $X$ or class space $C$. 
	Based on the above characterization, we then show how different properties of $Q$ affect 
	classification accuracy and generalization, such as the dimension of the Hilbert 
	space, the amount of noise, and the amount of neglected information from $X$ via, e.g., pooling layers.  
	Moreover, we introduce a quantum version of the Information Bottleneck principle that allows us
	to explore the various tradeoffs between accuracy and generalization. Finally,
	in order to check our theoretical predictions, we study the classification of the 
	quantum phases of an Ising spin chain, and we propose the 
	Variational Quantum Information Bottleneck (VQIB) method to optimize quantum embeddings of 
	classical data to favor generalization. 
\end{abstract}

\maketitle

\section{Introduction}

Quantum information and machine learning are two very active areas of research which 
have become increasingly interconnected \cite{biamonte2017quantum,schuld2018supervised,dunjko2018machine,carleo2019machine}.
In this panorama, many works have considered and designed learning models that are built by using quantum states and algorithms~\cite{schuld2019quantum,havlivcek2019supervised,lloyd2020quantum,perez2020data,cong2019quantum,ciliberto2020fast,huang2020power,banchi2020quantum,banchi2016quantum,gray2018machine,huang2021information,mitarai2018quantum,benedetti2019parameterized,liu2018differentiable,gambs2008quantum}.
Some of these proposals have focused on learning classical data by exploiting the
capability of quantum machines to easily perform computations that are 
in principle unfeasible using classical computers \cite{schuld2019quantum,havlivcek2019supervised}. 
Other works have instead focused on ``quantum data'', i.e., information embedded in quantum states or quantum channels.
 
For the latter case, a fundamental model with particular relevance is that of quantum channel discrimination.
This is known to have non-trivial implications for quantum sensing~\cite{sensingreview}, in tasks such as the detection of targets~\cite{QIll1,QIll2,QIll3} 
or the readout of memories~\cite{quantumreading1,quantumreading2}. Recently, it has been applied
to study the model of channel position finding~\cite{CPF}, associated with absorption spectroscopy~\cite{CPF2}, and more sophisticated problems of barcode decoding and pattern recognition~\cite{banchi2020quantum}.  In the latter,
the use of quantum light sources was shown to drastically reduce the error affecting the supervised classification of images, 
even when the output measurements are not optimized. 


The main difference between (supervised) quantum machine learning (QML) 
and quantum  hypothesis testing (QHT), such as state and channel discrimination \cite{bae2015quantum}, is the role of prior information. In QHT all possible sets of states and their prior probabilities are known. This is not the case in any machine learning approach, where instead prior information is in the form of {\it samples}, that is a collection of correctly classified states. These samples are not enough to cover all possible cases and the most important question in data driven strategies is to check for generalization:
\textit{After having trained the model using a few known examples,
can the model accurately classify unseen data?}
In previous QML literature, 
the generalization capabilities were numerically verified by 
computing the classification error over a testing set. 
Some theoretical bounds were studied in \cite{ciliberto2020fast,huang2020power,cheng2015learnability,abbas2020power}, but only for particular classifiers or regression models, while intuitive geometrical characterizations were given in \cite{lloyd2020quantum,perez2020data}, yet without a formal proof. 

Here we study generalization in QML classification tasks using tools from quantum information theory. In Sec.~\ref{s:background} we first establish a fruitful link between QML and QHT. This  allows us 
to bound the QML classification error by exploiting some QHT
results, 
and to study the role of unknown prior in QHT. 
In Sec.~\ref{s:main} we introduce the main technical result of this 
paper: quantities linked to either the training or testing errors can be
bounded by the quantum mutual information between some suitable quantum
states and classical variables.
Based on the study of these quantities, 
in Sec.~\ref{s:tradeoff} we show different implications of our theoretical
bounds: we introduce a quantum version of the bias-variance tradeoff,
which defines fundamental limitations on the testing error for finite amounts of data; 
we then show how to use results developed in the quantum communication/cryptography literature to study how to optimally embed classical information onto quantum states; 
finally, we show  how different properties 
of the quantum states affect the
classification accuracy and generalization, such as the dimension of the Hilbert 
space, the amount of noise, and the amount of neglected information via, e.g., pooling layers.  
Our results are based on the study of the linear loss function, yet we show how similar 
conclusions can be obtained in a loss-independent framework, by defining a quantum version of the information
bottleneck principle \cite{tishby2000information}.
In Sec.~\ref{s:appl} we consider different applications of our theoretical results. We first 
study the Quantum Phase Recognition problem of an exactly solvable quantum Ising chain and 
then define the Variational Quantum Information Bottleneck method to train quantum embeddings 
of classical data for good generalization. 
Conclusions are drawn in Sec.~\ref{s:conc}.
The mathematical derivations of our results, as well as the extension to
multi-ary classification, are presented in the appendices. 


\section{Quantum Hypothesis Testing vs Supervised Classification}
\label{s:background}

\begin{figure}[t]
	\centering
	\includegraphics[width=0.99\linewidth]{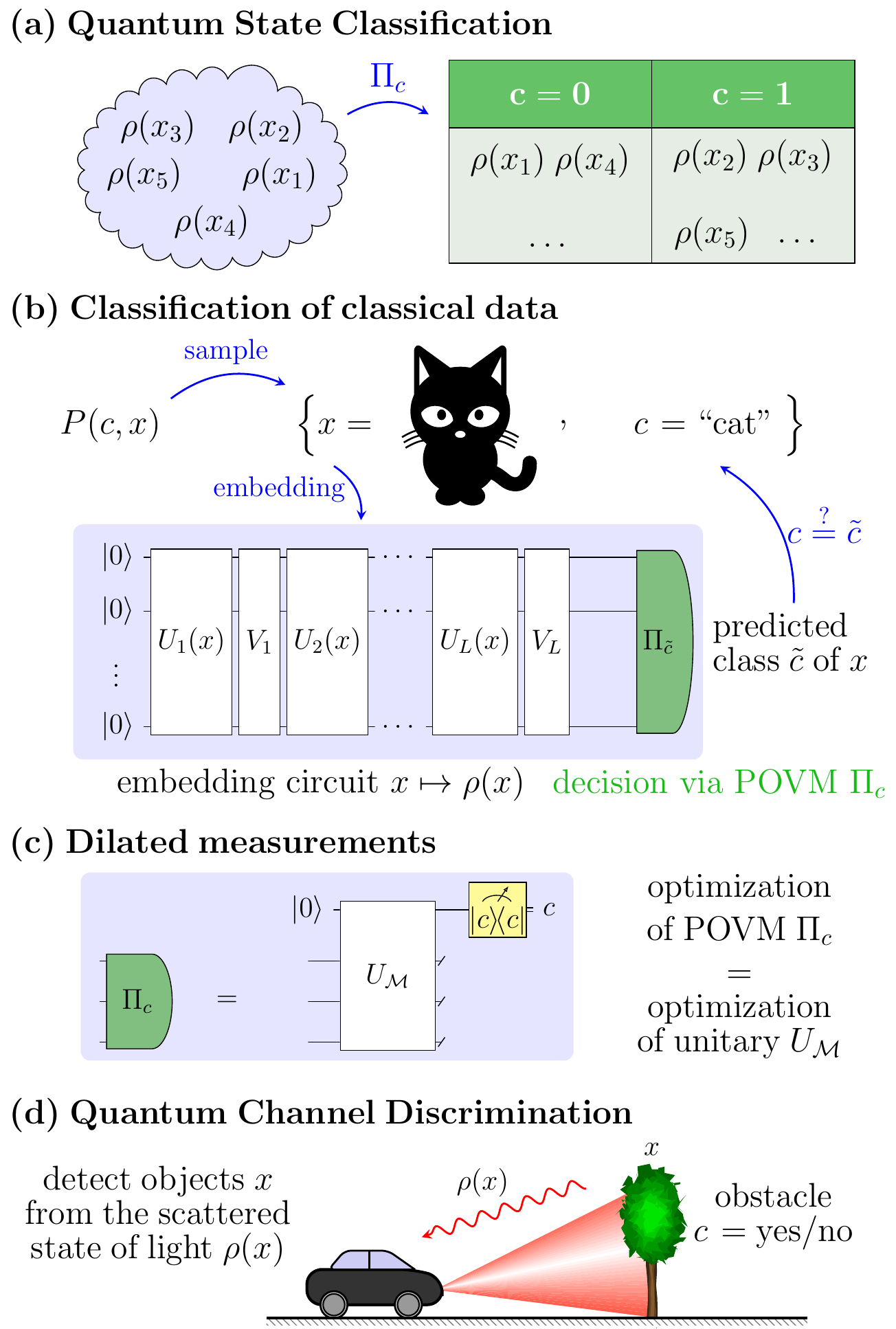}
	\caption{
		(a) Example binary classification of quantum states that depend on some external parameters $x$. 
		(b) Classification of classical data using a quantum embedding circuit with $L$ layers. The classical data
		are sampled from an {\it unknown} distribution $P(c,x)$, where $x$ describes, e.g., images of animals 
		and $c$ specifies the kind of animal, e.g., a cat.  The  classical
		input $x$ is embedded into a quantum state $\rho(x)$
		via layers of $x$-dependent and $x$-independent gates. A POVM $\{\Pi_c\}$ is performed 
		at the end of the circuit. The predicted class of $x$ corresponds to the measurement outcome $c$.
		(c) Any POVM can be expressed as a unitary circuit followed by a projective measurement $\ket{c}\!\bra{c}$ on 
		a suitably large ancillary system. 
		(d) In quantum channel discrimination, the images $x$ live in the physical
		world; a quantum probe senses the outside 
		world and $\rho(x)$ is the scattered state of light collected by the
		detector, which depends on the outside objects. The detector then classifies the image 
		with a POVM, as in (c).
	}%
	\label{fig:scheme}
\end{figure}

We study the classification of either quantum states, as in Fig.~\ref{fig:scheme}a, or
classical data, as in Fig.~\ref{fig:scheme}b, using the framework of QHT. Let us first 
consider the simpler case where a quantum 
device can only be in $N_C$ possible states $\{\rho_c\}_{c=1,\dots,N_C}$, for some integer $N_C$. 
The possible values of $c$ are called hypotheses in QHT, or {\it classes} in this paper. 
An experimentalist (Alice) performs a measurement on 
the device with a positive operator-valued measurement (POVM) $\{\Pi_c\}$, whose
outcome is the predicted value of $c$. 
Via Naimark's dilation theorem,
such a POVM can be effectively implemented as shown in  Fig.~\ref{fig:scheme}c,
namely by using an ancillary system whose Hilbert space dimension
is equal to $N_C$,
 by first applying a unitary circuit that couples the state and the ancilla, and then 
performing  a projective measurement $\ket{c}\!\bra{c}$ on the ancillary system. 
In the most general setting, the states
$\rho_c$ are not orthogonal and Alice cannot discriminate between them with a
single measurement. When the device can be 
reinitialized in the same state, Alice can use $N$ copies $\rho_c^{\otimes N}$ and 
the probability of wrong discrimination can decrease exponentially with $N$ \cite{montanaro2019pretty}. 
We remark that the common approach of using $N$ measurement ``shots'' is just a particular case, possibly 
non-optimal, of the above general framework, with independent measurements on each copy. 

Unlike QHT, in QML classification 
tasks there are different states that belong to the same class. The number of classes $N_C$ is 
finite, but the available states $\rho(x)$ are possibly infinite. In this paper the {\it inputs} $x$ model tunable classical 
parameters. For instance, consider a device that depending on parameters $x$ outputs either 
entangled ($c=1$) or separable ($c=0$) states \cite{lu2018separability}, 
or a many-body system that may be in different phases $c$ 
depending on some external magnetic fields $x$ \cite{cong2019quantum}. 
We may also be interested in classifying classical data $x$
(e.g.~images) using a quantum algorithm, to look for {\it algorithmic} quantum
advantage \cite{huang2020power,liu2020rigorous} (faster classification), or classifying quantum
channels using quantum probes to look for quantum advantage in 
accuracy \cite{banchi2020quantum} (fewer measurements).
When dealing with classical inputs $x$, the quantum embedding circuit can be written as in Fig.~\ref{fig:scheme}b 
with $x$-dependent and $x$-independent gates $U_i(x)$ and $V_j$, 
which may be optimized during training. 
Finally, a
mathematically related, yet physically different problem consists in classifying 
{\it physical} objects using quantum sensors
\cite{banchi2020quantum,QIll1,QIll2,QIll3,quantumreading1,quantumreading2,CPF,CPF2},
as in the example shown in Fig.~\ref{fig:scheme}d.
There, $\rho(x) = \mathcal E_x[\rho_{\rm in}]$ describes the state received by a quantum detector,
where $\rho_{\rm in}$ is the input probe state of light, possibly entangled, and 
$\mathcal E_x$ describes how the photons are scattered depending on the objects $x$ living 
in the physical world. 

In all the examples described above, 
we are interested in learning the unknown functional relation  
 $c=f(x)$ between a classical input $x$ and output class $c$, yet through measurements on a quantum device.
The motivations can be quite diverse and range from quantum device characterization depending 
on external parameters to the use of quantum algorithms to classify classical data. 
Following common practices in theoretical machine learning, we 
 assume that all possible pairs of data $(c,x)$ follow some unknown probability distribution 
 $P(c,x)$, so data pairs are independent samples from $P$ (see Fig.~\ref{fig:scheme}b).
Formally, our ignorance can be modelled using mixed states
\begin{equation}
	\rho_c = \sum_x P(x|c) \rho(x),
	\label{e:rhoc}
\end{equation}
where $P(x|c)$ is the unknown conditional probability. 
For $N$ copies, such states read 
$\rho_c^{(N)} = \sum_x P(x|c) \rho(x)^{\otimes N}$.
With a slight abuse of notation, to simplify the mathematical expressions we may hide the
dependence on $N$ inside $\rho(x)$, namely as 
$\rho(x) = \tilde \rho(x)^{\otimes N}$ for some $\tilde \rho(x)$. 
The main difference between QHT and the classification problem studied in this paper is that 
we take measurements on the instances $\rho(x)$ rather than onto the discrete states $\rho_c$. 
Measurements are still described via a POVM $\{\Pi_c\}$, possibly acting on $N$ copies, constructed such that its 
outcome $c$ is the predicted class $x$. 
Such a quantum classifier is probabilistic: given an input $x$ the predicted class $c$ is found 
with probability $p_{Q}(c|x) = \Tr[\Pi_c \rho(x)]$. Other classifiers can be built
using different techniques of quantum decision theory \cite{bae2015quantum,gambs2008quantum}, for instance by repeating 
the measurement many times and taking the most likely class, or 
by defining an observable $\mathcal M = \sum_c m_c \Pi_c$, for certain real
numbers $m_c$, and then assigning a certain class
depending on the expectation value $\Tr[\mathcal M \rho(x)]$. For instance, for binary classification 
problems with $c=\{0,1\}$ we may set $m_0=-1$, $m_1=1$ and then assign the class depending 
on the sign of  $\Tr[\mathcal M \rho(x)]$ \cite{lloyd2020quantum}. 
We remark that exact expectation values on real hardware can only be obtained in the limit of 
infinitely many shots, namely for $N\to\infty$ copies.

\begin{figure}[t]
	\centering
	\includegraphics[width=0.83\linewidth]{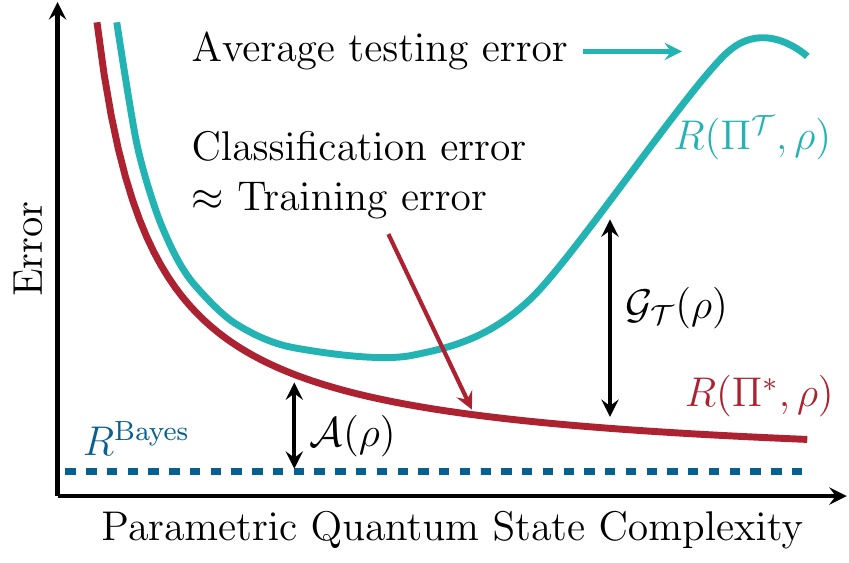}
	\caption{Summary of the error sources for given parametric quantum states $\rho(x)$ 
		and a finite number of training samples. 
		The classification error $R(\Pi^*)$ is the average loss with  the 
		unknown optimal measurement $\Pi^*$. The average testing error $R(\Pi^{\mathcal T})$ 
		replaces $\Pi^*$ with the POVM $\Pi^{\mathcal T}$ estimated from the training set $\mathcal T$ 
		via empirical risk minimization. The testing error $R^{\mathcal T'}(\Pi^{\mathcal T})$ 
		is a finite sample approximation of $R(\Pi^{\mathcal T})$ over the testing set $\mathcal T'$.
		The difference between the average testing error and the Bayes risk $R^{\rm Bayes}$ is split into 
		the approximation error $\mathcal A(\rho)$ and the generalization error $\mathcal G^{\mathcal T}(\rho)$. 
		The training error typically behaves as $R(\Pi^*)$.
	}%
	\label{fig:complexity}
\end{figure}

\begin{figure}[t]
	\centering
	\includegraphics[width=0.9\linewidth]{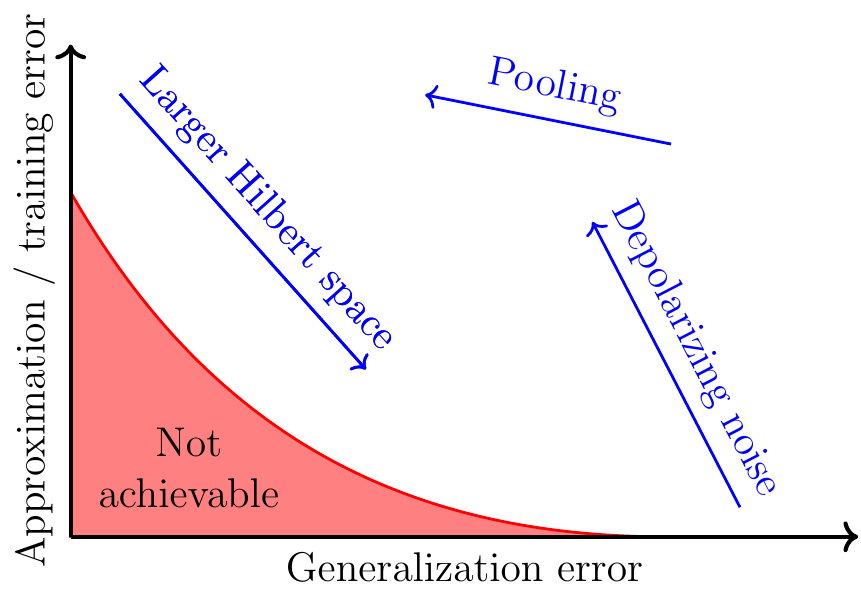}
	\caption{Summary of some of the main conclusions of this paper. The approximation and 
		generalization errors are mathematically related, respectively, to the training 
		and testing error over some datasets, and they cannot be simultaneously minimized (bias-variance tradeoff). 
		We use quantum information quantities to bound these errors and show how they are affected by the
		dimensionality of the quantum Hilbert space, noise, and ``information pooling''.
	}
	\label{fig:spoiler}
\end{figure}

Since in QML the probability distribution is unknown, a (sub)optimal classifier must be built 
from a finite amount of training data. Is the trained classifier able to
predict the correct class of previously unseen data? To answer this question,
in the next sections we use tools from quantum information theory to formally
study the two main errors, namely the approximation 
and generalization errors, which rigorously formalize the empirical testing
error (see Fig.~\ref{fig:complexity}).  We
call $\rho(x)$ parametric quantum states (PQS) that depend on some tunable
classical parameters $x$ and 
refer to the mapping $x\mapsto \rho(x)$ as {\it quantum embedding}.
We will study how different properties of the embedding  affect accuracy or generalization,
as schematically shown in Fig.~\ref{fig:spoiler}, and then introduce
fundamental limitations on the errors 
that we may expect for a given data distribution and finite training samples. 

\subsection*{Training and testing with linear loss}


We first formalize 
the various sources of error that may prevent generalization. 
Readers already familiar with this topic may skip this section and refer to Fig.~\ref{fig:complexity} 
for the notation. 

In supervised learning the available data are split between a training and a testing set. 
Both these sets are composed of pairs $(c_k,x_k)$, namely inputs $x_k$ and their 
true class $c_k$, but are used differently. 
We consider a training set $\mathcal T = \{(c_k,x_k)\}_{k=1,\dots,T}$ with $T$ pairs,
and similarly a testing set $\mathcal T'$ with $T'$ pairs.
In the training part a model
is optimized in order to minimize a suitable distance between the true class $c$ and 
the predicted class for all possible pairs $(c,x)\in\mathcal T$ in
the training set. For given PQS $\rho(x)$ and POVM $\{\Pi_c\}$ 
the quantum model predicts a class $\tilde c$ with probability $\Tr[\Pi_{\tilde c}\rho(x)]$, as in
Fig.~\ref{fig:scheme}. 
If $c$ is the true class of $x$, the linear {\it loss} is defined as the probability of 
misclassification, namely the probability that the predicted class $\tilde c$ is 
different from the true class $c$
\begin{equation}
	\ell(c,x) = \sum_{\tilde c\neq c} \Tr[\Pi_{\tilde c} \rho(x)] = 1-\Tr[\Pi_{c}\rho(x)],
	\label{e:qloss}
\end{equation}
where the second equality follows from $\sum_c \Pi_c=\openone$. 
The linear loss allows us to link QML to QHT \cite{bae2015quantum,gambs2008quantum}.

Training is done via empirical risk minimization, where the empirical risk 
is the average loss over all possible pairs $(c_k,x_k)$ in the training set
\begin{align}
	R^{\mathcal T}(\Pi) &= \frac1T\sum_{(c_k,x_k) \in \mathcal T} \ell(c_k,x_k), 
	\label{e:emplossPi}
\end{align}
and the minimization is over the parameters of the model, namely the POVM and,
in some applications, also the embedding.
In general such minimization does not have an analytic solution, except for a few notable 
cases. For binary classification problems, where  $c=\{0,1\}$ can only 
take two distinct values, 
the optimal $\mathcal T$-dependent POVM, 
$\Pi^{\mathcal T} = \argmin_\Pi [R^{\mathcal T}(\Pi,\rho)]$,
is the Helstrom measurement 
\cite{helstrom1969quantum}, which is extensively used in QHT.
The Helstrom measurement operator $\Pi_0^{\mathcal T}$ ($\Pi_1^{\mathcal T}$) 
is the projection onto the eigenspace of positive (negative) eigenvalues of 
$\frac{T_0}{T}\rho^{\mathcal T}_0 -\frac{T_1}T\rho^{\mathcal T}_1$, where 
$T_c$ is the number of inputs in the training set with class $c$ and 
$\rho_c^{\mathcal T}$
is a mixture of all the states $\rho(x)$ with inputs in $\mathcal T$ 
and fixed class $c$. Although not necessary, to simplify the equations 
we will always assume that the training set contains an equal number of 
inputs per class, so $T_c/T=1/2$. 
Using the optimal Helstrom measurement the minimum empirical risk can be written analytically 
in terms of the trace distance between the two average states $\rho^{\mathcal T}_0$ and  $\rho^{\mathcal T}_1$
\begin{equation}
	R^{\mathcal T}(\Pi^{\mathcal T}) = \frac12\left(1-\frac12\|\rho^{\mathcal T}_0 -\rho^{\mathcal T}_1\|_1\right)~.
	\label{e:trainerr}
\end{equation}
The above quantity is what defines the {\it training error} for 
a given PQS $\rho(x)$, namely the average loss 
over the training set. 
 From the above equation, zero training error is possible only when 
$\|\rho^{\mathcal T}_0 -\rho^{\mathcal T}_1\|_1=2$, which happens when 
$\rho^{\mathcal T}_0$ and  $\rho^{\mathcal T}_1$ have orthogonal support. 
We will show in Appendix~\ref{s:extended} that similar conclusions also hold when
the number of classes is greater than two. 

Does the model generalize? 
Empirically we need to check how the model performs with inputs not present in the training set. 
This is normally done by studying the testing error $R^{\mathcal T'}(\Pi^{\mathcal T})$, 
which is similar to Eq.~\eqref{e:emplossPi},
but where the samples are taken from the testing set $\mathcal T'$ and the POVM $\Pi^{\mathcal T}$ 
is the one minimizing the empirical risk. 
In order to define the generalization error more formally, 
we need first to define the true average classification error
\begin{equation}
	R(\Pi) = \ave_{(c,x)\sim P}  \ell(c,x) = 1-\sum_c P(c) \Tr[\Pi_c\rho_c],
	\label{e:qrisk}
\end{equation}
where in the last expression we use the chain rule $P(c,x)=P(x|c)P(c)=P(c|x)P(x)$,
$P(x) = \sum_c P(c,x)$, $P(c)=\sum_x P(c,x)$ 
and the definition of Eq.~\eqref{e:rhoc}. 
The training error \eqref{e:emplossPi} is an empirical approximation of the
classification error \eqref{e:qrisk}
where the formal average over all possible pairs $(c,x)$ is substituted with a finite average 
over the training set. The optimal classification POVM, 
$\Pi^*=\argmin_\Pi R(\Pi)$, is in
general different from the $\Pi^{\mathcal T}$ that we get from empirical 
risk minimization. Overfitting happens when this difference is significant, namely when 
$\Pi^*$ and $\Pi^{\mathcal T}$ disagree on the class of an input not present in the training set. 
The generalization error, also called the {\it estimation error}, is defined as 
$	R(\Pi^{\mathcal T})-R(\Pi^*) $, namely as a difference
between two classification errors, where in one case we use the true classifier and in the other we use the classifier built 
from the training set $\mathcal T$. 
Notice that the testing error $R^{\mathcal T'}(\Pi^{\mathcal T})$ 
is an empirical approximation of $	R(\Pi^{\mathcal T})$.

In order to have a low testing error, we need to have both a low
generalization error and a low classification error $R(\Pi^*)$.
The lowest possible classification error is the (normally unknown) Bayes
classifier $R^{\rm Bayes}$ -- see next section for a formal definition. The 
difference between the average testing error $R(\Pi^{\mathcal T})$ 
and $R^{\rm Bayes}$ can be written as 
\begin{equation}
	R(\Pi^{\mathcal T}) - R^{\rm Bayes} = \mathcal G^{\mathcal T}(\rho) + \mathcal A(\rho)~,
	\label{e:RAG}
\end{equation}
with 
\begin{align}
	G^{\mathcal T}(\rho) &= R(\Pi^{\mathcal T})-R(\Pi^*),
	\label{e:Gdef} \\
	\mathcal A(\rho) &= R(\Pi^*) - R^{\rm Bayes}.
	\label{e:Adef0}
\end{align}
In Eq.~\eqref{e:RAG} the difference between the average testing error
and the Bayes risk has been 
split into two positive terms (see also Fig.~\ref{fig:complexity}): 
$\mathcal G^{\mathcal T}(\rho)$ is the previously defined generalization error while 
$\mathcal A(\rho)$ is called the {\it approximation error}. 
A standard result of statistical learning theory, dubbed the bias-variance tradeoff \cite{shalev2014understanding}, 
shows that it is impossible to minimize both $\mathcal A$ and $\mathcal G$. 
 Simple classifiers may escape from overfitting but have a bias in the
resulting predictions, while too complex classifiers lead to
overfitting and a higher variance in the predictions. 
We remark that these complexity analyses cannot explain the success of deep learning, 
where models with millions of parameters generalize well in spite of their complexity. 
There are some explanations of why deep-learning works in particular models \cite{belkin2019reconciling,canatar2021spectral}, but 
this is still a subject of intensive research. Moreover, quantum models that can be 
trained in near-term quantum hardware are quite far from the regime where 
deep learning operates, so in this paper we focus on models of ``moderate complexity''. 

\section{Quantum information bounds for supervised learning}\label{s:main}

In this section we study bounds on the approximation and generalization errors using 
tools from quantum information theory, the main theoretical results of this paper.
In Sec.~\ref{s:tradeoff} we study how different
properties of the PQS $\rho(x)$ affect these errors, while in Sec.~\ref{s:appl} 
we study more practical applications.

\subsection{Generalization error}

Employing tools from statistical learning theory \cite{shalev2014understanding} 
and quantum information, in Appendix~\ref{s:extended}
we prove one of our main results:

\begin{thm}\label{t:main}
For a given embedding $x\mapsto \rho(x)$ and for any $\delta>0$,
with probability at least $1-\delta$, the generalization error is bounded as
\begin{equation}
	G^{\mathcal T} 
	\leq
2\sqrt{\frac{\mathcal B}{T}} + \sqrt{\frac{2\log(1/\delta)}{T}},
	\label{e:genboundPi}
\end{equation}
where $T$ is the size of the training set,
\begin{equation}
	\mathcal B = \left(\Tr\sqrt{\sum_x P(x) \rho(x)^2}\right)^2=2^{I_{2}(X:Q)}
	\label{e:Bdef}
\end{equation}
depends on the embedding, and $P(x)$ is the (unknown) prior probability for an
image $x$. 
\end{thm}

We refer to $\mathcal B$ as the {\it generalization bound}, which 
constrains how large the generalization error can be for a fixed number $T$ of 
training pairs. 
The inequality \eqref{e:genboundPi} applies to binary 
classification problems, but its general form, derived  in Appendix~\ref{s:extended}, 
is equivalent to \eqref{e:genboundPi} up to a constant that depends on the 
number of (equiprobable) classes -- see Theorem~\ref{t:p2}. 
The inequality \eqref{e:genboundPi},  with the explicit form of $\mathcal B$ in \eqref{e:Bdef},
represents one of the central results of this paper, as it links the generalization 
error to properties of the embedding that are measured by information theoretic 
quantities. Indeed, the quantity found in the second equality of Eq.~\eqref{e:Bdef} 
is the 2-Renyi mutual information between subsystems $X$ and $Q$ of the classical-quantum state
\begin{equation}
	\rho_{CXQ} = \sum_{cx} P(c,x)~ \ket{cx}\!\bra{cx} \otimes \rho(x).
	\label{e:rhoextended}
\end{equation}
For general $\alpha$ and subsystems $A$ and $B$, the $\alpha$-Renyi mutual information 
\cite{berta2015renyi} is defined as 
\begin{align}
	I_{\alpha}(A{:}B) &= \frac\alpha{\alpha-1} \log_2\Tr
	\sqrt[\alpha]{
	\Tr_A \left(\rho_{A}^{\frac{1-\alpha}2} \rho_{AB}^\alpha
\rho_{A}^{\frac{1-\alpha}2} \right)}. 
\label{e:IalphaAB}
\end{align}
For $\alpha\to1$ one recovers the quantum mutual information 
$I_1(A{:}B) \equiv I(A{:}B) = H(A) + H(B)-H(AB)$, where $H(A) = S(\rho_A)$ and $S(\rho) = -\Tr[\rho\log_2\rho]$
is the von Neumann entropy. 
Although $I_{\alpha}(A{:}B) $ for $\alpha\neq 1$ does not satisfy all of the properties of $I_1(A{:}B) $, it does 
satisfy the data processing inequality \cite{berta2015renyi}, namely $I_{\alpha}(A{:}B) \geq I_{\alpha}(A'{:}B')$ under local quantum channels $\mathcal E^{A/B\to A'/B'}$, a central 
ingredient in quantum information theory. 

In Eq.~\eqref{e:rhoextended} 
we have introduced three Hilbert spaces: the quantum space $Q$ where the PQSs $\rho(x)$ live,
the class space $C$ spanned by  $\{\ket{c}\}_{c=1,\dots,N_C}$ and the input space $X$ spanned by 
$\{\ket{x}\}$ for all possible values of $x$, namely where each input $x$
is mapped onto a different orthogonal state $\ket x$. 
For instance, if the inputs $x$ are made of classical images with $n$ pixels, each with 
a 16 bit color, then $\ket{x}$ lives in a space of ${4n}$ qubits. 
For continuous inputs, e.g.~when $\rho(x)$ is an equilibrium state of a
many-body system and $x$ some external parameters, 
one must consider a suitably-regularized infinite dimensional Hilbert space. Here 
for simplicity, we assume that $X$ is discrete and can be represented using $N_X$ classical bits,
so $\rho_{CXQ}$ lives in a $N_C2^{N_X+N_Q}$ dimensional Hilbert space.

Inequalities as in \eqref{e:genboundPi} are common in statistical learning theory and 
show that, with high probability, a model generalizes well whenever $T\to\infty$. 
The importance of \eqref{e:Bdef} is in quantifying when the size $T$ of the training set is ``large''. 
According to our analysis, a training set is large whenever $T\gg 2^{I_2(X{:}Q)}$, namely 
when $\log_2(T)$ is much larger than the number of bits required to describe the information 
shared between the input distribution and the quantum embedding, as measured by $I_2(X{:}Q)$.

\subsection{Approximation error}

Fixing the embedding $x\mapsto\rho(x)$ is like fixing the model class in classical machine learning, 
e.g.~a neural network with a given architecture and a certain number of nodes. 
The difference between the minimum classification error with a given architecture 
and the theoretical minimum over all possible architectures, namely the Bayes risk, 
is the approximation error \eqref{e:Adef0}. 
For a known $P(c,x)$ and a given $x$, the Bayes classifier 
picks the class that maximizes $P(c|x)$. The corresponding Bayes 
risk for binary classification problems with $P(c)=1/2$ is  then
\begin{align}
	R^{\rm Bayes} &= 1-\frac12 \sum_x \max\{P(x|0),P(x|1)\}= \frac{1-\Delta}2,
	\nonumber 
\end{align}
where 
\begin{equation}
	\Delta = \frac12
	\sum_x \big|P(x|0)-P(x|1)\big|.
\end{equation}
Using the definition of the approximation error \eqref{e:Adef0} and 
the classification error $R(\Pi^*)$, which is analogous to \eqref{e:trainerr} but 
with the states \eqref{e:rhoc}, we find that the approximation error for 
quantum binary classification problems can be written as 
\begin{align}
	\mathcal A  &= R(\Pi^*) - R^{\rm Bayes} = \label{e:Adef}
	 \Delta
							- \frac{\|\rho_0-\rho_1\|_1  }2.
\end{align}
It is simple to show that $0 \leq \mathcal A \leq \Delta$. Indeed, 
the upper bound is trivial, and can be achieved when 
$\rho_0=\rho_1$. As for the lower bound, by defining $\rho_c^{XQ} = \sum_x P(x|c) \ket{x}\!\bra{x}
\otimes \rho(x)$, we first note by explicit calculation that $ \|\rho_0^{XQ}-\rho_1^{XQ}\|_1=2\Delta$.
Then by the contractivity of the trace distance over quantum channels we find 
$ \|\rho_0^{Q}-\rho_1^{Q}\|_1  \leq \|\rho_0^{XQ}-\rho_1^{XQ}\|_1 $,
where $\rho^Q_c=\Tr_X [\rho_c^{QX}]\equiv  \rho_c$, thus showing that $\mathcal A\geq 0$.
The approximation error $\mathcal A$ can be interpreted as a generalization of the probability of error
in QHT, where the difference is due to the measurements over the instances $\rho(x)$ rather than 
over the discrete hypothesis states \eqref{e:rhoc}. The two errors coincide up to a multiplicative factor 
when $X\equiv C$.


In the previous section, we have showed how to bound the generalization error using the mutual 
information between subsystems  $X$ and $Q$ in \eqref{e:rhoextended}. 
We can use entropies to bound the average classification error $R(\Pi^*)$,
and hence $\mathcal A$. Indeed, 
using the quantum Chernoff bound \cite{audenaert2007discriminating} and explicit 
calculations we get 
$R(\Pi^*) \leq \frac12 \Tr[\sqrt{\rho_0}\sqrt{\rho_1}] = {2^{-I_{\frac12}(C{:}Q)}}-\frac12$, which 
shows that a low classification error is possible when $I_{\frac12}(C{:}Q)$ is large. A more general result, 
valid for any number of classes, can be found using conditional entropies \cite{konig2009operational},
\begin{equation}
	R(\Pi^*) = 1 - 2^{-H_{\rm min}(C|Q)} \leq 1- \frac{2^{I(C{:}Q)}}{N_C},
	\label{e:riskIQC}
\end{equation}
which is valid for states of the form given in Eq.~\eqref{e:rhoextended}, where $H_{\rm min}(C|Q) \leq H(C|Q)$ is 
the min- conditional entropy,  which is smaller than von Neumann's conditional entropy 
\cite{tomamichel2009fully}. The second inequality comes from $H(C|Q) = H(C)-I(C{:}Q)$ 
and $H(C)=\log_2 N_C$ for a classification problem with $N_C$ classes. Since $Q$ and $C$ 
are classically correlated, $I_1(C{:}Q)\leq H(C)$  and 
a small risk is possible when the mutual information between $Q$ and $C$ is large. 

To conclude, small $\mathcal G$ is possible for small $I_2(X{:}Q)$ while small $\mathcal A$ is 
possible for large $I(C{:}Q)$. 
In the following section, we build upon these theoretical bounds to study how different properties 
of the PQS $\rho(x)$ affect the approximation and generalization errors. 


\section{Bias variance tradeoff for quantum machine learning}\label{s:tradeoff}

The bias-variance tradeoff is a central result in machine learning,
stating that it is impossible to minimize both the approximation and the generalization errors. 
Models with lots of parameters and structure are expected to have low approximation error, potentially at the 
cost of poor generalization (overfitting). 
On the other hand, a low-dimensional model with few parameters would be easier
to learn, but it might not reliably classify the data.

We study the bias-variance tradeoff using quantum information. 
Remember that the difference 
between the average testing error and the Bayes risk can be written as a sum of two positive terms 
\eqref{e:RAG}, the generalization error $\mathcal G$ and the approximation error $\mathcal A$. 
The latter is the difference between the average classification error $R$ and the Bayes risk,
while the former can be bounded by the generalization bound $\mathcal B$ from \eqref{e:Bdef},
and can be made arbitrarily small by considering arbitrarily large training sets with $T\to\infty$. 
For finite and fixed $T$, we show that it is impossible to minimize 
both $\mathcal A$ and $\mathcal G$, 
and that different properties of PQS $\rho(x)$ affect the 
approximation and generalization errors, as schematically shown in Fig.~\ref{fig:spoiler}. 

Thanks to our framework, many characterizations of PQS $\rho(x)$ will be formally derived
in the next section using tools from quantum information. 
We will use, in particular,
the contractivity of the trace distance 
under quantum channels $\mathcal E^{Q\to Q'}$ 
\cite{watrous2018theory},
mapping from the space $Q$ to the space $Q'$, i.e., $\|\mathcal E^{Q\to Q'}(\rho-\sigma)\|_1
\leq \|\rho-\sigma\|_1$, 
the data processing inequality \cite{berta2015renyi}  $I_2(X{:}Q)\geq I_2(X{:}Q')$,
and finally the bounds satisfied by $\mathcal B$, studied in the Appendices~\ref{s:extended}
and \ref{s:further}
\begin{equation}
	1 \leq \mathcal B \leq 2^{\min\{H_2(X),N_Q\}},
	\label{e:Bbounds}
\end{equation}
where $D=2^{N_Q}$ is the dimension of the embedding Hilbert space, i.e.~$N_Q$ is the 
number of qubits in the PQS, and 
$H_\alpha[X] = \frac\alpha{\alpha-1}\log_2[\sum_x P(x)^{\frac1\alpha}]$, 
is the R\'enyi entropy of the classical input distribution.

\subsection{Properties of quantum embeddings}

Here we focus on the mapping $x\mapsto \rho(x)$ and discuss some properties and 
desirable features.

\emph{ It is impossible to minimize both $\mathcal B$ and $R$}:
The optimal embedding is the one that {discards} all the {\it irrelevant} information from the input space $X$ that 
is not necessary to predict the class $C$. Indeed, according to Eqs.~\eqref{e:Bdef} and 
\eqref{e:riskIQC}, $I_2(X{:}Q)$ must be small while $I(C{:}Q)$ must be large. 
We now show that it is impossible to minimize both $\mathcal B$ and $R$ by studying 
the two extreme cases where the information about $X$ is either fully discarded or fully maintained. 

\emph{ Constant embeddings provide zero generalization error, but the largest
approximation error}: indeed,
the generalization error \eqref{e:Gdef} is defined as the distance between the 
risk obtained by minimizing the empirical loss over the training data and the true average loss.
For constant embeddings this difference is zero: if we restrict ourselves to classifiers 
that always produce a constant answer, then it is trivial to learn this classifier from 
data, but the average classification error will be as high as 50\%.
Mathematically, from the definition \eqref{e:Bdef} and the bounds \eqref{e:Bbounds},
it is clear that the minimum $\mathcal B$ 
is achieved with the trivial constant embedding, $\rho(x) = \rho$ for all $x$,
but such a trivial embedding provides both the largest $R=1/2$ and the largest 
training error $R^{\mathcal T} = 1/2$ 
from Eq.~\eqref{e:trainerr}. Moreover, using mutual informations, the constant
embedding is the only embedding for which the
space $Q$ is uncorrelated from both $C$ and $ X$ 
in \eqref{e:rhoextended} and so $I(X{:}Q)=I(C{:}Q)=0$.

\emph{ Basis encoding guarantees zero approximation error, but the largest generalization error}. Basis 
encoding \cite{farhi2018classification,schuld2018supervised} is defined as $\rho(x) = \ket{x}\!\bra{x}$, namely 
different inputs are mapped onto orthogonal vectors on a suitably large Hilbert space. 
No information is lost, or hidden, in the quantum embedding, so 
using Eqs.~\eqref{e:Adef} and \eqref{e:Bbounds}, we get the lowest possible approximation error $\mathcal A=0$,
meaning the average loss can achieve the Bayes risk. However, for the same reason, we also get 
the largest $\mathcal B=2^{H_2(X)}$, since $X=Q$. Therefore, basis encoding allows us to reach the theoretical minimum
classification error, but it requires a large embedding space and many ($T\gg \mathcal B$) training pairs 
to avoid overfitting. 

\begin{figure}[t]
	\centering
	\includegraphics[width=0.8\linewidth]{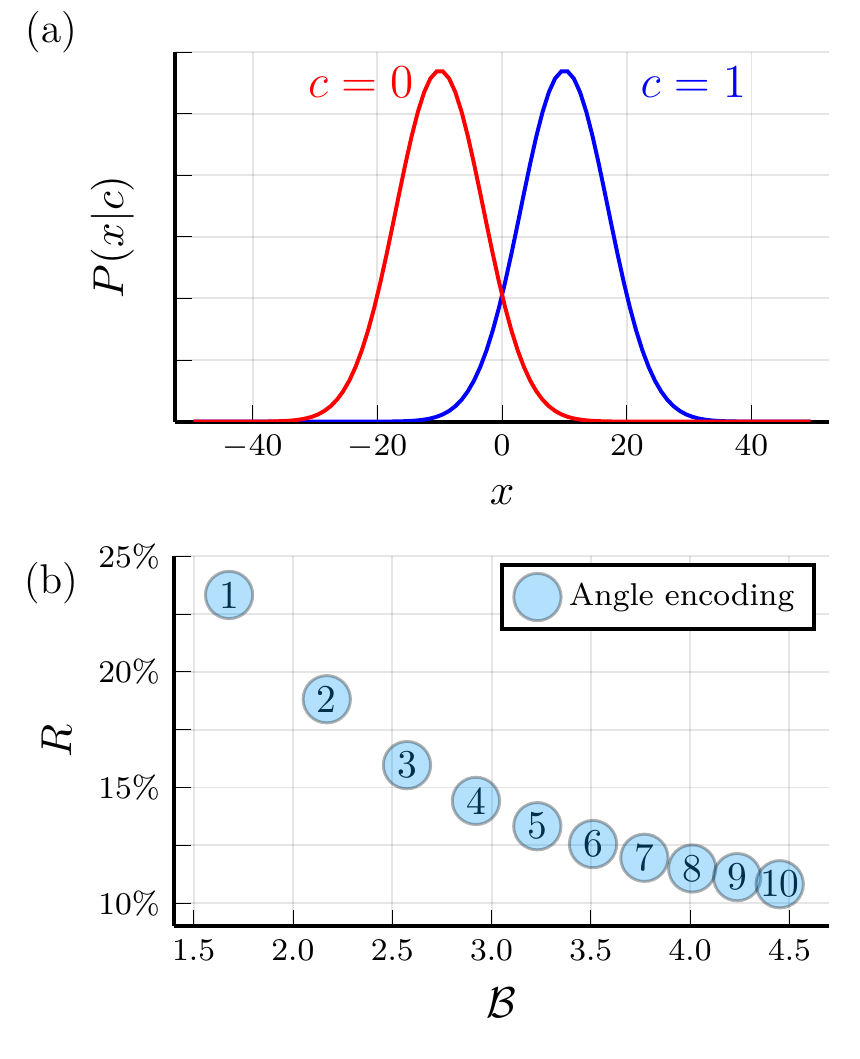}
	\caption{(a) Example data distributions for two different classes 
		$c=\{0,1\}$: Gaussian distributions with different means $\pm10$ and the same 
		standard deviation $7$. The corresponding Bayes risk is $R^{\rm Bayes}\simeq 7.6\%$. 
		(b) Average classification error $R$ \eqref{e:qrisk} vs generalization bound $\mathcal B$ \eqref{e:Bdef} 
		for the angle encoding $\rho(x) = \ket{\psi(x)}\!\bra{\psi(x)}^{\otimes N_Q}$ 
		where $\ket{\psi(x)} = \cos(x/2)\ket0+\sin(x/2)\ket 1$ and the $x$-values have been normalized 
		such that $x\in[0,2\pi]$. The number inside each circle represents the value of 
		$N_Q=1,\dots,10$. 
	}%
	\label{fig:angle}
\end{figure}

\emph{High-dimensional embeddings may have lower approximation error}: 
 in Appendix~\ref{s:extended} 
(Theorem~\ref{t:approx}) we show that if we define an embedding by taking $N$ copies 
of a simpler one, i.e.~if we consider $x\mapsto \rho(x)^{\otimes N}$, then 
$\mathcal A\to 0$ for $N\to \infty$ as long as $F(\rho(x),\rho(y))\neq 0$ for $x\neq y$,
where $F(\rho,\sigma)=\|\sqrt{\rho}\sqrt{\sigma}\|_1$ is the fidelity between two quantum 
states. 
Intuitively this happens because asymptotically it is possible to correctly discriminate all 
the states $\rho(x)$ via QHT \cite{montanaro2019pretty,banchi2020quantum}, effectively achieving 
a basis encoding for $N\to\infty$.
If $\rho(x)$ is an $N_Q$ qubit state, then optimized embeddings using $N\times N_Q$ 
qubits can only have a lower approximation error than $\rho(x)^{\otimes N}$, since the 
latter is a particular case. 
However, high dimensional embeddings may suffer from poor generalization, as $\mathcal B$ 
may be larger. 
A numerical check of this prediction is shown in Fig.~\ref{fig:angle} for a binary 
discrimination problem with Gaussian priors. We see that, as the number of qubits 
increases, the classification error risk quickly decreases but the generalization bound  increases. 
This is consistent with our numerical prediction. According to our Theorem~\ref{t:approx}, 
for many-copies, $\mathcal A$ decreases exponentially with $N$. The asymptotic behaviour 
of $I(X{:}Q)$ is still an open question, but we note that for local measurements $I(X{:}Q)$ can be 
bounded by the mutual information between an input random variable and $N$ output observations, 
which may display two different regimes $\mathcal O(\log N)$ or $\mathcal O(N)$ \cite{haussler1997mutual}. 
Therefore we conjecture that $\mathcal B$ may slowly increase with $N$ (e.g.~polynomially) for 
particular datasets and embeddings, as we observe numerically for small $N$.


\emph{ Low-entropy datasets and low-dimensional embeddings can in principle
generalize well}: this is a trivial consequence 
of Eq.~\eqref{e:Bbounds}, when the entropy of the dataset is measured by $H_2[X]$. 
The statement about the dimension can be made a bit more precise by focusing not just on 
the dimension of the Hilbert space, but on how much the information is distributed within
the Hilbert space. For instance, let us assume a pure state embedding $\rho(x) = U(x)\ket0\!\bra0 U(x)^\dagger$
with a unitary embedding circuit $U(x)$. 
If the embedding is such that the input information is ``fully-scrambled'' in a $d$-dimensional 
subspace, with $d\ll 2^{N_Q}$, then we may write $\sum_x P(x) \rho(x)^2 \approx \openone_d/d$. Substituting this 
approximation in \eqref{e:Bdef} we get 
\begin{equation}
	\mathcal B \approx \mathcal O(d)~.
\end{equation}
Therefore, an embedding is capable of generalization not just when built using few qubits, but rather when
it ``scrambles'' information in a small subspace of the full $N_Q$-qubit Hilbert space.

\begin{figure}[t]
	\centering
	\includegraphics[width=0.5\linewidth]{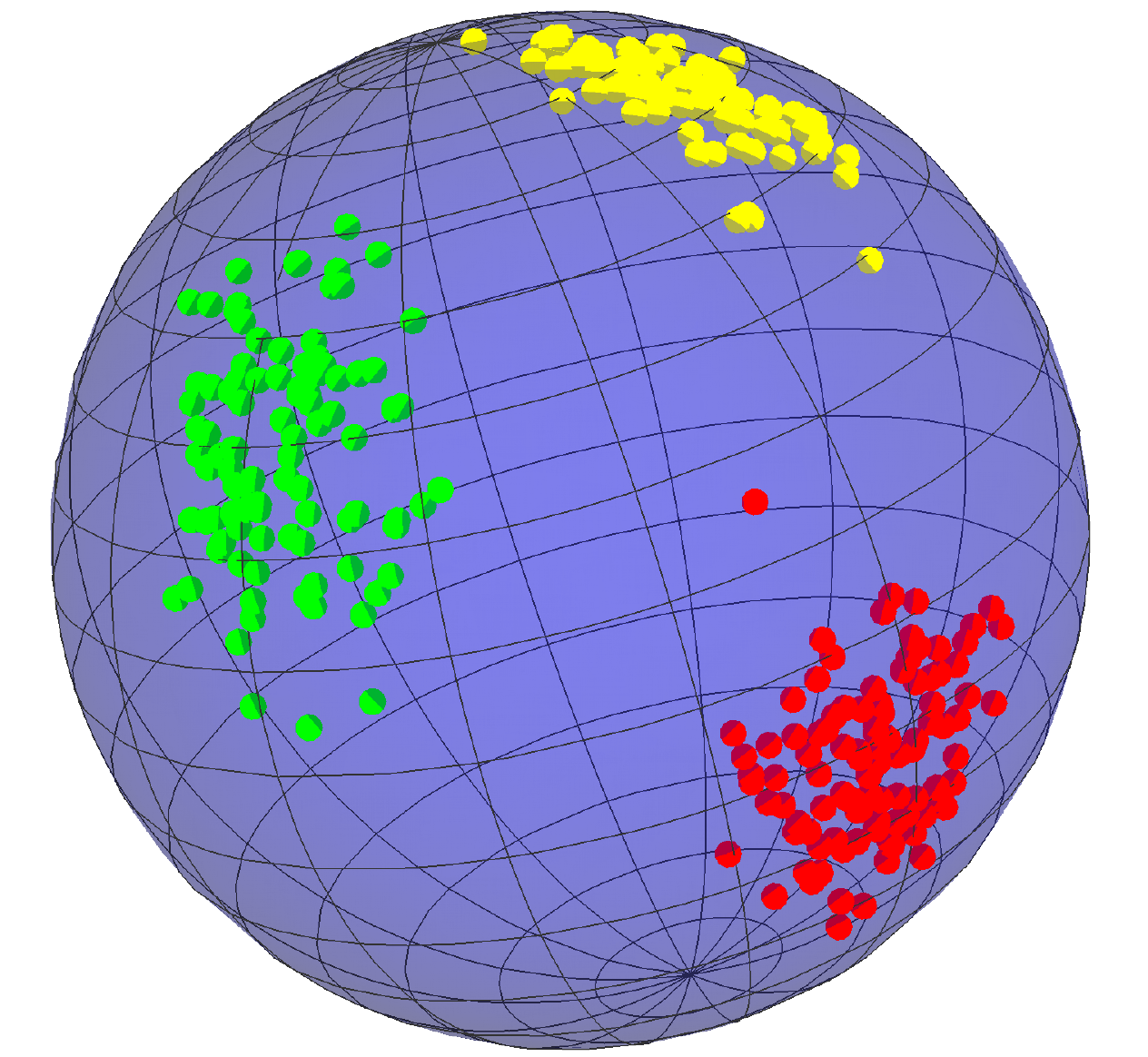}
	\caption{Geometric visualization of a good embedding, where each point in the sphere represents 
		a state. Points belonging to different classes are plotted with different colors. 
		A good embedding should cluster points with the same class and separate points belonging to different classes. 
	}%
	\label{fig:sphere}
\end{figure}

\emph{ Geometric characterization}: There is an intuitive geometrical characterization 
of ``good'' embeddings (see e.g.~Ref.~\cite{perez2020data,lloyd2020quantum}). 
A good embedding is possible when the fidelity 
between two embedded states is small if the inputs are from different classes and high if the inputs 
are from the same class, as schematically shown in Fig.~\ref{fig:sphere}. This intuitive picture 
can be explicitly proved using our results.  Indeed, using 
the Fuchs-van de Graaf inequality and  the strong concavity of the fidelity,
we get \begin{equation}
	\mathbb E^{\mathcal T}  R^{\mathcal T}(\Pi^{\mathcal T}) \leq \frac12 \mathbb{E}^{\mathcal T}
	\left[F(\rho_0^{\mathcal T},\rho_1^{\mathcal T})\right]
	\leq \frac12 F(\rho_0,\rho_1),
	\label{e:averisk}
\end{equation}
so, on average, low training errors and low classification errors are possible when the fidelity between 
the two averaged states $\rho_0$ and $\rho_1$ is small. Moreover, 
in Appendix~\eqref{s:further} we show that $\mathcal B\leq (\sum_c \sqrt{P(c)\mathcal B_c})^2$ and,
for pure state embeddings, 
\begin{equation}
	\mathcal B_c \leq 1 + 
						\sqrt{(r_c^2-r_c)\left(1 -\Tr[\rho_c^2]\right)}.
	\label{e:boundpurity}
\end{equation}
where $r_c$ is the rank of $\rho_c$. The general case is discussed 
in Appendix~\eqref{s:further}.
The above inequality 
shows that low generalization error is possible when the average embedding states $\rho_c$ have 
low rank and/or high purity. Since $\rho_c$ is an ensemble of embeddings for inputs from the same class, 
the above requirement is satisfied when $\rho(x)$ effectively maps all the
inputs from the same class to the same state. 
More precisely, for pure state embeddings 
$\rho(x) = U(x)\ket0\!\bra0 U(x)^\dagger$, the purity can be written as 
\begin{equation}
	\Tr[\rho_{c}^2] = \sum_{x,y} P(x|c) P(y|c) F(\rho(x),\rho(y))^2,
	\label{e:purity}
\end{equation}
where $F(\rho(x),\rho(y))= |\bra0 U(y)^\dagger U(x)\ket 0|$ is the fidelity.
Therefore, good generalization is possible whenever $F(\rho(x),\rho(y))$  is large 
 for all possible pairs $(x,y)$ of inputs with the same class, namely when
$\rho(x)$ and $\rho(y)$ are always geometrically close in the embedding Hilbert space. 
Combining this with \eqref{e:averisk} we see that a desirable feature to get a good embedding 
is that 
the fidelity between two embedded states is small if the inputs are from
different classes and high if the inputs are from the same class, as in 
Fig.~\ref{fig:sphere}.

\emph{ Noisy operations}: We focus on what happens when 
$\rho(x) = (1-\epsilon) U(x)\ket0\!\bra0 U(x)^\dagger + \epsilon \openone/2^{N_Q}$, namely
when the embedding discussed in the previous example 
is degraded by depolarising noise with strength $\epsilon$. 
Again assuming that the average 
fully scrambles information in a $d$-dimensional subspace, then 
\begin{equation}
	\sum_x P(x) \rho(x)^2 \approx \left[(1-\epsilon)^2+\frac{2(1-\epsilon)\epsilon}{2^{N_Q}}\right]\frac{\openone_d}{d} 
	+ \frac{\epsilon^2}{2^{2N_Q}}\openone~,
	\nonumber
\end{equation}
from which 
\begin{equation}
	\mathcal B\approx \left(\sqrt{d}(1-\epsilon+\epsilon/2^{N_Q}) + (1- d/2^{N_Q})\epsilon\right)^2
	\simeq d(1-\epsilon)^2.
	\nonumber
\end{equation}
From the above equation, we see that the generalization error does not increase
with noise. It actually decreases when $1\ll d \ll 2^{N_Q}$, 
as for large $\epsilon$ the embedding approaches the constant embedding, which 
has the lowest generalization error but the highest classification and approximation errors. 

\emph{Kernels close to identity are not good for generalization}: quantum embeddings 
can be used to define quantum kernels \cite{schuld2019quantum,havlivcek2019supervised,lloyd2020quantum,huang2020power}. 
These ``kernels'' are nothing but the fidelity between two states. 
Sometimes working with kernels rather than quantum states can be beneficial. In Appendix~\ref{s:kernels} we show that 
for pure state embeddings $\rho(x)=\ket{\psi(x)}\!\bra{\psi(x)}$
we can express the generalization bound quantity as 
\begin{align}
	\mathcal B = (\Tr\sqrt K)^2,
	\label{e:kernel}
\end{align}
where $K$ is the normalized kernel operator, whose matrix elements are  
\begin{equation}
	K_{x,y} = p(x)\langle\psi(x)\ket{\psi(y)}.
\end{equation}
Accordingly, the generalization bound can be computed from the eigenvalues $\eta_k$ of the normalized 
kernel matrix as $\mathcal B = (\sum_k \sqrt{\eta_k})^2$. Note that the study of normalized kernel eigenvalues 
was also recently proposed as a possible explanation of the generalization
capabilities of deep learning \cite{canatar2021spectral}. 
From the above expression, one readily finds that worse generalization
performances, according to 
our bounds \eqref{e:genboundPi} and \eqref{e:Bbounds} are obtained when 
$\langle\psi(x)\ket{\psi(y)} \simeq \delta_{x,y}$, namely when different states have almost orthogonal support. 
Indeed, when $\langle\psi(x)\ket{\psi(y)}=\delta_{x,y}$ Eq.~\eqref{e:kernel} results in the upper bound of 
\eqref{e:Bbounds}.

\emph{Pooling may help}: We have seen that a large embedding Hilbert space may favour the classification accuracy,
yet hinder generalization. According to \eqref{e:Bbounds} the generalization 
bound may approach its largest value when $N_Q$, i.e., the number of qubits in the embedding, 
is as large as $H_2(X)$. What about the minimum number of qubits? For binary classification 
problems a good embedding can be obtained even with $N_Q=1$. Indeed,
the simplest embedding that achieves the minimum 
Bayes risk is $\rho^{\rm Bayes}(x) =\ket {\tilde c}\!\bra {\tilde c}$, where  $\tilde c=\argmax_c P(x|c)$ 
is, for a given $x$, the class with largest conditional probability. 
Clearly it is impossible to construct this embedding, as the probabilities $P(x|c)$ are unknown, but 
the above example shows that a good embedding is possible with a single qubit. 
Although the state before the measurement must be as low-dimensional as possible, 
we may start from a large dimensional embedding and then iteratively {\it throw away information}, either 
via measuring some qubits and then applying a different unitary on the remaining ones depending on the 
measurement result  or, equivalently, by applying a conditional gate and then discarding some qubits via a partial 
trace. 

Since the generalization error depends only on the dimension of the 
final Hilbert space, one can use pooling to iteratively reduce the number of qubits, using different 
layers, eventually leaving a single qubit for measurements. 
Promising forms of pooling have been proposed as a basis for Quantum Convolutional Neural Networks (QCNN)
\cite{cong2019quantum,oh2020tutorial}, where the pooling layers are constructed using a
reverse Multiscale-Entanglement-Renormalization-Ansatz (MERA) circuit,
whose depth depends logarithmically on the total number of qubits. 
QCNNs have some desirable features, such as the ability to distinguish states corresponding to 
complex phases of matter \cite{cong2019quantum}, and the lack of barren plateaus in their 
parameter landscape \cite{pesah2020absence}, which aids training. 
Our analysis shows that QCNNs, or other embeddings built by iteratively pooling information, 
also have good generalization capabilities.

\subsection{Quantum Information Bottleneck}\label{s:ib}

In the previous section we showed that it is impossible to minimize both the
approximation and the generalization errors, when these are defined starting from the linear loss 
\eqref{e:qloss}. We now show how the generalization/approximation tradeoff can also be understood
from information theoretic principles that are independent of the choice of loss function. 
 In  classical settings, a method designed for this purpose 
is the information bottleneck (IB) principle \cite{tishby2000information,fischer2020conditional}, whose aim is to 
find the ``best'' compressed representation $Z$ of the input $X$ that nonetheless has all 
the relevant information required to predict the class $C$. The amount of compression can be quantified 
using the classical mutual information $I(X{:}Z)$, while $I(C{:}Z)$ quantifies the residual information between $C$ and $Z$. In order to have accurate classification 
$I(C{:}Z)$ must be large, while to have good compression $I(X{:}Z)$ must be small. 
The information bottleneck principle finds a compromise between accuracy and compression by 
minimizing the Lagrangian $\mathcal L_{IB} = I(X{:}Z) - \beta I(C{:}Z)$ for a certain value of $\beta$.
The parameter $\beta$ allows us to {\it explore} different regimes and to favour either 
accuracy or compression. 
When $\beta=0$ the minimization of $\mathcal L_{IB}$ achieves the best compression, without caring 
about correct classification. While for $\beta\to\infty$ the minimization of  $\mathcal L_{IB}$
achieves optimal classification without compression.  Quantum generalizations 
of the information bottleneck principle were considered for quantum communication problems in 
\cite{salek2018quantum,datta2019convexity}. Here we apply the IB principle to the different 
problem of finding the optimal embedding.  

We focus on the state \eqref{e:rhoextended} where $X$ and $C$ are, respectively, the classical spaces 
of inputs and classes, and $Q$ is the quantum embedding Hilbert space. 
We then define the quantum IB Lagrangian as
\begin{equation}
	\mathcal L_{IB} = I(X{:}Q)-\beta I(C{:}Q),
	\label{e:ibL}
\end{equation}
where $I(A{:}B) = I_{\alpha\to 1}(A{:}B)$ in \eqref{e:IalphaAB} is the quantum mutual information.
Both $I(X{:}Q)$ and $I(C{:}Q)$,
can be expressed using Holevo's accessible information \cite{holevo1973bounds}.
In \eqref{e:genboundPi}, \eqref{e:Bdef} and \eqref{e:riskIQC}, we have shown that good generalization is possible 
whenever $I_2(X{:}Q)$ is small, while low classification error is possible when $I(C{:}Q)$ is large. 
These conclusions were found for the linear loss \eqref{e:qrisk}. We may 
assume that $I_\alpha(X{:}Q)$ defines a family of generalization bounds for different loss functions, 
so the minimization of the generalization error is consistent with the minimization of $I(X{:}Q)$, 
according to some metric, while the maximization of $I(C{:}Q)$ is consistent with the accurate prediction 
of $C$ from $Q$. For a particular value of $\beta$, the optimal embedding is then obtained as 
$	\min_{\rho(x)} \mathcal L_{IB}. $
From the definition, we find the explicit form of the IB Lagrangian as 
\begin{align}
	\mathcal L_{IB} =& \nonumber 
	(1-\beta) S\left[\sum_x P(x) \rho(x)\right] - \sum_x P(x) S[\rho(x)]+ \\
									 &+ \beta\sum_c P(c) S[\rho_c] + \sum_x \tilde\lambda_x \Tr[\rho(x)]+\eta~,
									 \label{e:iblext}
\end{align}
where $S[\rho]$ is the von Neumann entropy of $\rho$, $\tilde\lambda_x$ are Lagrange multipliers 
to force correct normalization, and 
$\eta$ contains all the terms that are independent of the embedding. 
The optimal embedding corresponds to a minimum of $\mathcal L_{IB}$, which satisfies 
$\frac{\partial \mathcal L_{IB}}{\partial \rho(z)} =0$. By explicit computation 
we find that the above condition defines a recursive equation for the optimal embedding
\begin{align}
	\tilde\lambda_z	\rho(z) = e^{(1-\beta)\log \bar \rho 
	+ \beta\sum_c P(c|z) \log \rho_c},
	\label{e:ibsolmixed}
\end{align}
where $\bar \rho = \sum_c P(c) \rho_c$ and $\lambda_z$ is directly related to $\tilde\lambda_z$ and 
is needed to enforce normalization. Alternatively, by restricting to pure state embeddings 
$\rho(x) = \ket{\psi(x)}\!\bra{\psi(x)}$, we get 
\begin{equation}
	\tilde\lambda_z	\ket{\psi(z)} = e^{(1-\beta)\log \bar \rho 
	+ \beta\sum_c P(c|z) \log \rho_c}\ket{\psi(z)}.
	\label{e:ibsolpure}
\end{equation}
From Eqs.~\eqref{e:ibsolmixed} and \eqref{e:ibsolpure} we see that, for $\beta=0$, we get 
a constant embedding, while for large $\beta$ the optimal embedding for a given $x$ 
is iteratively obtained from one of the  eigenvectors of $\sum_c P(c|x) \log \rho_c$ with the largest 
eigenvalue, or a mixture of them.

\begin{figure}[t]
	\centering
	\includegraphics[width=0.99\linewidth]{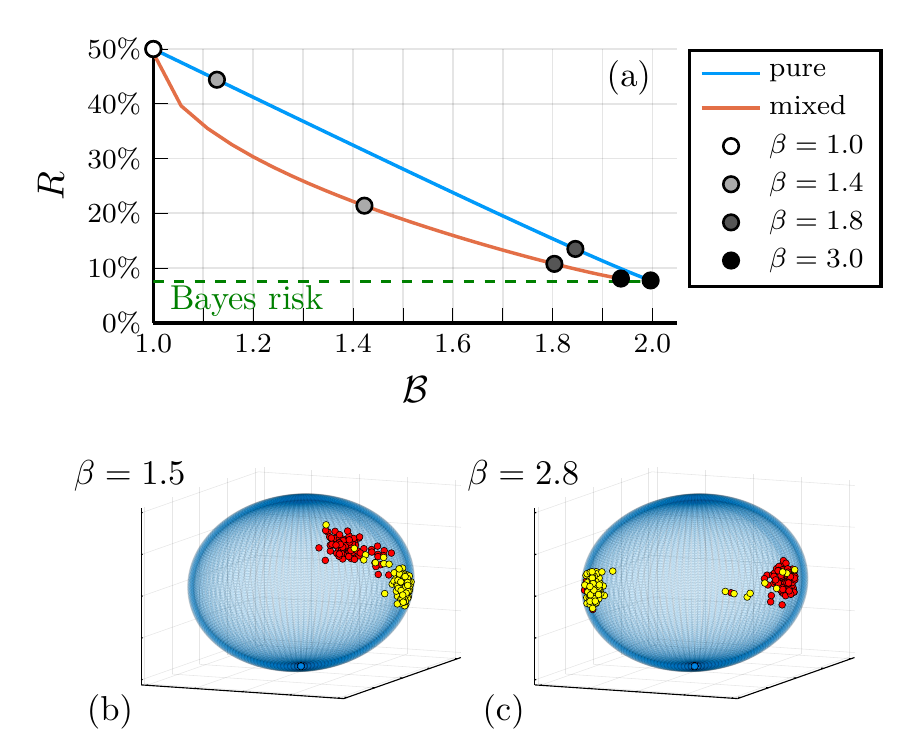}
	\caption{ (a) Average risk (approximation error) $R$ \eqref{e:qrisk} vs.
		generalization error $\mathcal B$ \eqref{e:Bdef} 
		for the optimal embeddings obtained by solving the IB equations using either pure 
		\eqref{e:ibsolpure} or mixed \eqref{e:ibsolmixed} single-qubit states for
		different values of $\beta\in[1,3]$.
		Some values of $R$ and $\mathcal B$ for particular $\beta$ are also shown with markers. 
		(b) and (c) Bloch sphere visualization of the quantum embeddings $\rho(x)$ for 
		different values of $x$ sampled from either $P(x|0)$ (red) or $P(x|1)$ (yellow), obtained 
		by solving the IB equations \eqref{e:ibsolpure} for single-qubit encodings, using the two 
		shown values of $\beta$. 
		In all three subfigures, the data distributions are those of Fig.~\ref{fig:angle}(a). 
	}
	\label{fig:ib}
\end{figure}

A numerical solution of the IB equations is shown in Fig.~\ref{fig:ib}, where ``pure'' and ``mixed'' 
refer to either \eqref{e:ibsolpure} or \eqref{e:ibsolmixed}, which 
were solved for two Gaussian distributions and a single-qubit embedding,
using a fixed number of iterations (1000). The Bayes risk is plotted as a reference, giving the smallest 
possible value of $R$ that can be obtained with any embedding. 
For a fixed $1\leq \beta\leq 3$, we first compute 
the optimal embedding via either \eqref{e:ibsolpure} or \eqref{e:ibsolmixed}, and then 
compute the classification error $R$ from Eq.~\eqref{e:qrisk}
and the generalization bound $\mathcal B$ from Eq.~\eqref{e:Bdef}. Recall that,
for a given $P(c,x)$,
the classification error is equivalent to the approximation 
error $\mathcal A$, up to the constant Bayes risk \eqref{e:Adef0}.
Fig.~\ref{fig:ib}(a) shows the approximation-generalization tradeoff for the 
different regimes that we have explored by varying $\beta$. For low values of $\beta$, we get 
an almost constant embedding with a large classification error (up to 50\%) and low $\mathcal B$, while 
for large values of $\beta\geq 2$, we find that $R$ approaches the theoretical lower bound (Bayes risk),
but at the expense of a larger generalization error, as $\mathcal B$ gets close to the theoretical upper bound 
\eqref{e:Bbounds}. We point out though that for this particular 
example, with a single-qubit embedding and two Gaussian priors, the
generalization error is always low due to the bound 
\eqref{e:Bbounds}.

The properties of the optimal embedding are shown in Figs.~\ref{fig:ib}(b) and (c). 
In particular, in panel (b) we observe that data belonging to different classes are clustered, but 
not well separated from each other. On the other hand,  for larger values of $\beta$, points belonging to different classes 
are typically very far apart in the Bloch sphere, though there are still some points in the wrong cluster. 
This prediction is consistent with the analysis of the fidelity between two different embeddings discussed 
in the previous section and sketched in Fig.~\ref{fig:sphere}: a good embedding is such that $F(\rho(x),\rho(y))$ 
is large whenever $x$ and $y$ belong to the same class and small otherwise.

\section{Applications}\label{s:appl}

In this section we study two different applications of our theoretical results. 
The first one deals with ``quantum data'', where the parametric quantum states $\rho(x)$ 
are fixed by the problem. The second one focuses on the classification of classical data, where 
the quantum embedding $x\mapsto\rho(x)$ can be optimized. In this latter case, we propose 
the Variational Quantum Information Bottleneck (VQIB) method for optimizing embeddings 
in order to favour generalization. 

\subsection{Quantum Phase Recognition}\label{s:qpr}

In Quantum Phase Recognition \cite{cong2019quantum} 
the task is to recognise the phases of matter of a quantum many-body system, 
by taking measurements on the quantum device itself, without having access to a 
classical description of its state. 
Here we focus on a paradigmatic  exactly solvable model of quantum statistical mechanics,
namely the one-dimensional transverse-field Ising model \cite{mbeng2020quantum}
\begin{equation}
	H = -\sum_{i=1}^L\left(\sigma^x_i \sigma_{i+1}^x + h \sigma_i^z\right),
	\label{e:Ising}
\end{equation}
where $\sigma_j^{x,y,z}$ are the Pauli matrices acting on site $j$ and we consider periodic boundary 
conditions, $\sigma_{L+1}^\alpha \equiv \sigma_1^\alpha$. 
For this model, the classical input is the magnetic field $h\equiv x$. 
In the thermodynamic limit $L\to\infty$,  the model displays a quantum phase
transition at the critical value $h=1$, separating an ordered phase for $|h|<1$
with two-fold degenerate ground states from a disordered phase for $|h|>1$ with unique ground 
state. The model can be exactly solved via fermionization \cite{mbeng2020quantum}. To simplify our analysis 
for finite $L$, here we ignore the subtleties of the different fermion parity sectors by considering a
small symmetry-breaking term that that forces the ground state to have even
parity. In that case, for even $L$ the ground state can be expressed as \cite{zanardi2007ground}
\begin{equation}
	\ket{\Phi_{\rm GS}(h)} = \bigotimes_{k=1}^{L/2} \left(
	\cos(\theta_{k,h}/2)\ket{00}_{k} + 
	\sin(\theta_{k,h}/2)\ket{11}_{k}\right),
\end{equation}
where $\ket{00}_{k}$ and $\ket{11}_{k}$ are respectively the vacuum and occupied states by two fermion 
pairs with opposite momentum $k,-k$, and 
\begin{align}
	\theta_{k,h} &= \arccos\left(\frac{c_k-h}{1+h^2-2hc_k}\right), &
c_k = \cos\frac{2\pi k}L.
\end{align}
From the above expression, it is trivial to compute the overlap
$f(h,h') = \langle{\Phi_{\rm GS}(h')} \ket{\Phi_{\rm GS}(h)} = \prod_{k=1}^{L/2} \cos\left(\frac{\theta_{k,h}-\theta_{k,h'}}2\right)$.
\begin{figure}[t]
	\centering
	\includegraphics[width=\linewidth]{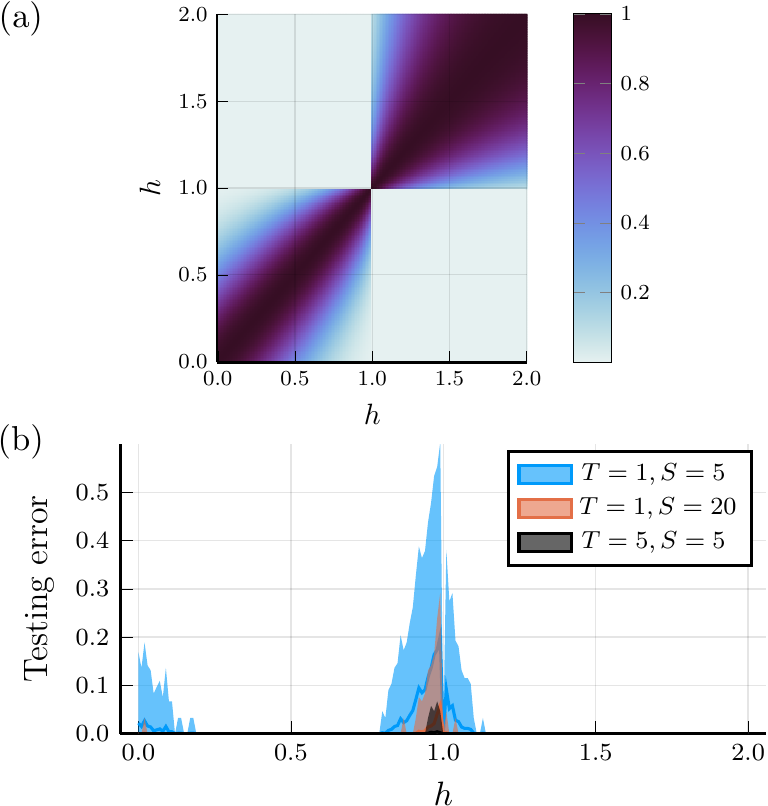}
	\caption{(a) Fidelity between two ground states of the quantum Ising model
		with different values of the magnetic field $h$, for $L=100$. The model displays a quantum phase transition at the 
		critical value $h=1$, separating ordered ($|h|<1$) and disordered ($|h|>1$) phases.  
		(b) Testing error in Quantum Phase Recognition 
		as a function of the magnetic field $h$. We use the fidelity classifier with a training set of $T$ random elements 
		per phase.
		Each fidelity is estimated via a SWAP test with $S$ shots. For each $h$ the fidelity is calculated $1000$ times. 
		Solid lines represent the mean fidelity, while shaded areas are the confidence intervals within a standard deviation. 
	}%
	\label{fig:ising}
\end{figure}
In the thermodynamic limit the fidelity induced distance $1-f(h,h+\epsilon)$ for small $\epsilon$ diverges at 
the critical point \cite{zanardi2007ground}. Therefore, we may expect that the fidelity between two states from the different phases 
become very small. This is indeed shown in Fig.~\ref{fig:ising}(a). Geometrically this means that the states belonging 
to different phases are clustered in distant areas of the Hilbert space, as in Fig.~\ref{fig:sphere}. However, 
$f(h,h')$ decreases exponentially in $L$ for $h\neq h'$, so for large $L$ the matrix $f(h,h')$ is almost diagonal,
thus signalling bad generalization performances according to our Eq.~\eqref{e:kernel}. 

A scaling analysis of $\mathcal B$ as a function of $L$ is beyond the scope of this work. In what follows we 
test our theoretical predictions for a fixed chain length $L=100$. In this case, we consider a uniform 
distribution $P(h)$ over $[0,2]$ and compute $\mathcal B$ from \eqref{e:kernel} -- where $x$ there is 
the magnetic field $h$. More specifically, we have discretized the interval such that \eqref{e:kernel} can 
be computed from the numerical eigenvalues, and we have observed that the result converges to $\mathcal B\simeq 5.9$ 
for 100 discretization points. 

We then train a fidelity classifier \cite{gambs2008quantum} to recognise the phases of the quantum Ising model 
\eqref{e:Ising}. In general the fidelity classifier associates 
to an unknown state $\ket\psi$ the class of the state from the training set with highest fidelity with $\ket\psi$. 
Such fidelity can be estimated via the SWAP test using $S$ shots, namely $S$ copies of $\ket\psi$. Since 
the swap test measurement operator is idempotent, the result of the SWAP test is a Bernoulli random variable with mean $F$, 
the fidelity, and variance $F(1-F)/S$. 
The fidelity measurement provides a non-optimal classification POVM, so this classifier 
is expected to perform slightly worse then the optimal strategies discussed theoretically 
in the previous sections. 

For numerical simulations 
we consider a training set with $T$ random elements with $h>1$ and $T$ 
random elements with $h<1$, and verify the Quantum Phase Recognition problem by generating new 
testing states $\ket{\Phi_{\rm GS}(h)}$ for $h$ uniformly distributed in $[0,2]$. 
In Fig.~\ref{fig:ising}(b) we numerically observe that even with $T=1$ the testing error is almost zero, 
except near the critical point. By increasing the number of shots, the fidelity is estimated more precisely, 
and given that states belonging to different phases have very low fidelity, as shown in Fig.~\ref{fig:ising}(a),
the testing error decreases. When $T \approx \mathcal B$ the training error is normally very low, except near 
the critical point. For $T=10\gg \mathcal B$ we always find zero training error, irrespective of the number of shots. 
Therefore, this analysis confirms the predictions of our Theorem~\ref{t:main}.

\subsection{Variational Quantum Information Bottleneck}\label{s:vqib}

We now focus on using a quantum algorithm to classify classical data. In this case, 
the states $\rho(x)$ are not fixed by the problem, as in the previous section, 
and can be optimized together with the measurement POVM. 
The embedding $x\mapsto\rho(x)$ can be optimized by training a 
quantum circuit as in Fig.~\ref{fig:scheme}. More specifically, we consider one of the simplest 
yet most general classification circuits with a single-qubit classifier, dubbed 
``data reuploading'' \cite{perez2020data}: here we use a slightly modified version where 
the embedding is obtained as a composition of $L$ layers of $x$-dependent 
single-qubit rotations around the $y$ and $z$ axes 
\begin{equation}
	\ket{\psi_w(x)} = \prod_{\ell=1}^L \left[
		R^z\left(w^{z\ell}\cdot x {+} w^{z\ell}_0\right)
		R^y\left(w^{y\ell}\cdot x {+} w^{y\ell}_0\right)
	\right]\ket 0,
	\label{e:reup}
\end{equation}
where $R^\alpha(\theta) = e^{i\theta\sigma^\alpha}$, $\sigma^\alpha$ are the Pauli matrices 
and the weight tensor $w^{\alpha\ell}_k$ can be optimized during training.

Based on the Quantum Information Bottleneck principle proposed in
Sec.~\ref{s:ib} we study the variational minimization of
the QIB Lagrangian \eqref{e:iblext} with respect to the 
parametric states \eqref{e:reup}. 
For single-qubit states, the entropies in Eq.~\ref{e:iblext} 
can be expressed without loss of generality 
in terms of the purity as 
\begin{equation}
	S(\rho) = 
	-(\lambda_- \log_2\lambda_-)
	-(\lambda_+ \log_2\lambda_+)
	=: s(\mathcal P),
\end{equation}
where 
\begin{equation}
	\lambda_\pm(\rho) = \frac{1\pm\sqrt{2\mathcal P(\rho)-1}}2,
\end{equation}
are the eigenvalues of $\rho$, which only depend on the purity 
$	\mathcal P(\rho) = \Tr[\rho^2] $. 
Since the state \eqref{e:reup} is pure, $S(\rho(x))=0$ in Eq.~\eqref{e:iblext}. 
Moreover, in order to train the embedding, we approximate the averages over 
the distribution $P(c,x)$ with empirical averages over 
the elements of the training set $\mathcal T$, so from 
Eq.~\eqref{e:iblext} we get 
\begin{equation}
	\mathcal L_{\rm IB}^{\mathcal T} = (1-\beta) s(\mathcal P_{\rm tot})
	+\beta \sum_c \frac{T_c}T s(\mathcal P_c),
	\label{e:vib}
\end{equation}
where constant terms have been neglected, and by explicit computation, the 
purities read
\begin{align}
\mathcal P_{\rm tot} &= \frac{T+2\sum^{\mathcal T}_{x < y} F(\rho(x),\rho(y))^2}{T^2}, \\
\mathcal P_{c} &= \frac{T_c+2\sum^{\mathcal T_c}_{x < y} F(\rho(x),\rho(y))^2}{T_c^2},
\end{align}
where 
 $\sum^{\mathcal T}$ refers to the double sum over the elements $(c_x,x), (c_y,y)$ 
from the training set, while in $\sum^{\mathcal T_c}$ the sum is restricted 
over elements with class $c_x=c_y=c$. The ordering $x<y$ refers to the index of 
the inputs in the training set, and is used just to avoid double counting.

\begin{figure}[t!]
	\centering
	\includegraphics[width=0.99\linewidth]{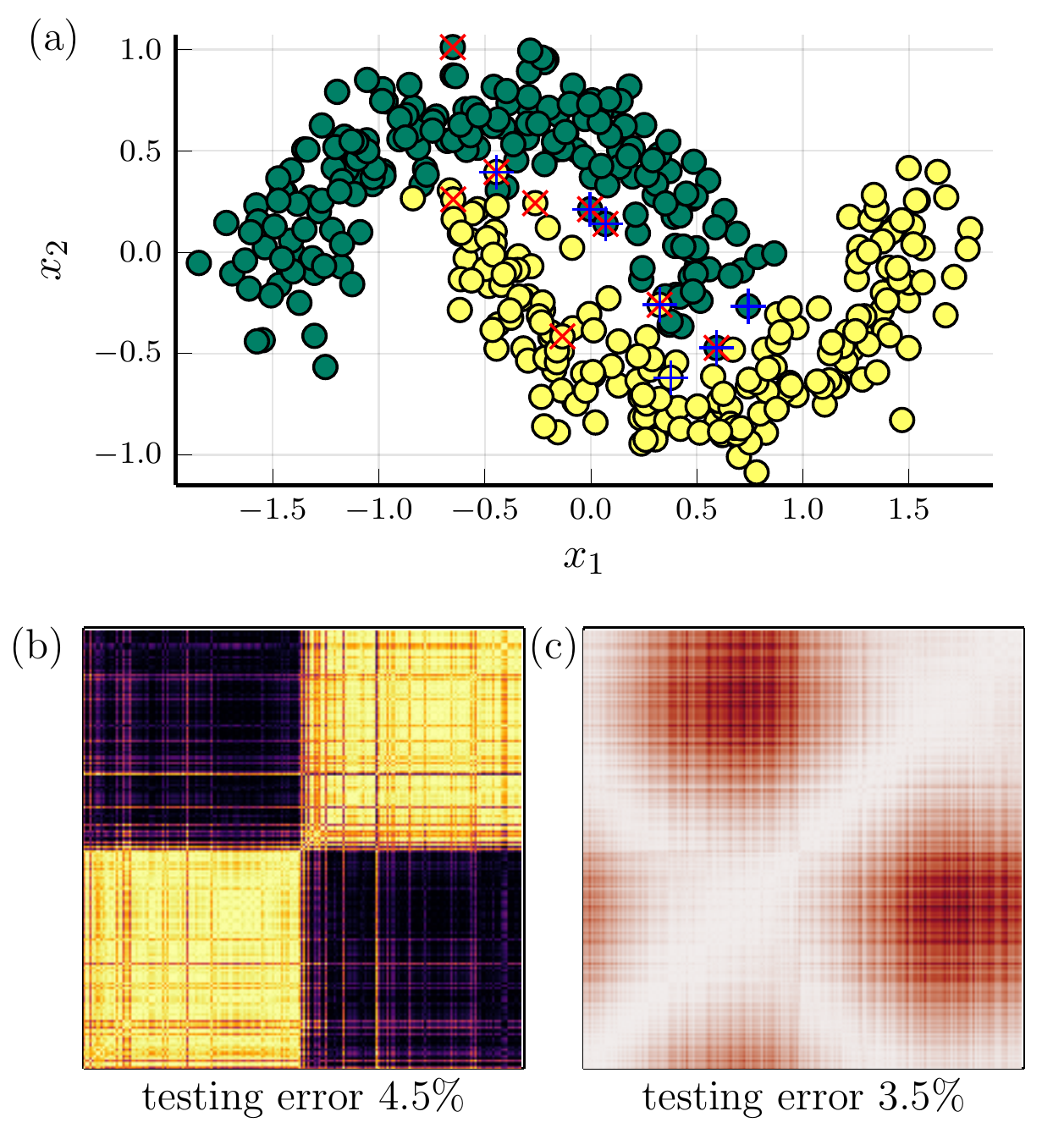}
	\caption{(a) Union of the training and testing sets from a generated 2-moon dataset. 
		Data (filled circles) with different colors belong 
		to two different classes. Wrongly classified data in the testing set after 
		training are marked with a red diagonal cross ($\beta=30$)
		or with a blue cross ($\beta=1.5$). 
		(b) Fidelity $F$ between two embeddings for $\beta=30$, using the data from (a). 
		Data points were ordered to first have all points from the first class and then all points from second class. 
		Dark black points represents $F\simeq 0$, while light yellow points represent $F\simeq 1$. 
		(b) Fidelity $F$ between two embeddings, as in (b) but for  $\beta=1.5$. White points represent 
		$F=1$ while dark red points have infidelity $1-F\simeq 10^{-7}$. 
	}%
	\label{fig:reup}
\end{figure}

As an example for numerical simulations, we
consider a binary classification problem with the  2-moons dataset shown in
Fig.~\ref{fig:reup}(a), where each point 
is described by two real coordinates $x\equiv (x_1,x_2)$. Moon points are organized 
in the two different patterns shown with different colors in Fig.~\eqref{fig:reup}(a),
which represent the two classes. Data have been generated using a noise parameter $0.3$, 
which makes the classification less deterministic. 
We generate a training set of 100 samples per class and 
optimize Eq.~\eqref{e:vib} using the Nelder-Mead algorithm with starting point
$w_k^{\alpha\ell}=0$ (constant embedding). In Figs.~\ref{fig:reup}(b) and (c), we show the fidelity between two 
trained embeddings $F(\rho(x),\rho(y))$, 
where training was performed using either $\beta=30$ or $\beta=1.5$.
After training, we use the fidelity classifier \cite{gambs2008quantum} 
to study both the training and testing errors. Unlike the previous section, here we study an 
exact evaluation of the fidelity, which would require an infinite amount of measurement shots. 
The training error we get with the optimized embedding is always zero. 
This is consistent with our theoretical analysis  (see Theorem~\ref{t:approx} in Appendix~\ref{s:extended}), 
as for $N\to\infty$ copies we may formally get zero approximation error. 

As shown in Figs.~\ref{fig:reup}(b), for large $\beta$ the trained embedding is able to separate 
most data points belonging to different classes into almost orthogonal quantum states. 
More precisely, the fidelity is 
almost zero for most inputs belonging to different classes, yet being mostly 
very high for states belonging to the same class, thus signalling good generalization. 
Indeed, by generating a testing set with 100 elements per class (also shown in 
Fig.~\ref{fig:reup}a), we observe a testing error $\simeq 4.5\%$. With a much
larger testing set of $10^4$ points we get a testing error of $\simeq 2.6\%$. 

Nonetheless, even better generalization can be obtained for $\beta=1.5$, although 
the optimized embedding is almost constant, as shown in Fig.~\ref{fig:reup}(c),
with largest infidelity $\simeq 10^{-7}$. The testing errors over the testing sets 
of 100 or $10^4$ elements per class described above are respectively $3.5\%$ and $1.9\%$, 
both smaller than those obtained with larger $\beta$. 
The price to pay is that, due to the small infidelities, many more measurements are needed 
to estimate the fidelity with the due high-precision for correct discrimination. 

The wrongly classified samples in the smaller testing set are shown in Fig.~\ref{fig:reup}(a) 
with a cross. We observe that for the small $\beta=1.5$ only the elements
near the boundaries may be 
wrongly classified, while for the larger $\beta=30$, in spite of neater class separation shown 
in Fig.~\ref{fig:reup}(b), there are wrongly classified samples in the ``bulk'' of the moons. 
Something similar was also observed in the numerical simulation shown in Fig.~\ref{fig:ib}(c).

Our analysis shows that the Variational Quantum Information Bottleneck method can be 
successfully used to train quantum embeddings with different generalization properties.

\section{Conclusions}\label{s:conc}
We have introduced measures of complexity to {\it quantify} 
the generalization and approximation capabilities QML classification problems, either with general parametric quantum states $\rho(x)$ or
quantum embeddings $x\mapsto \rho(x)$ of classical data $x$, 
when optimal measurements are performed on the system. 
One of the main results of this paper is the bound on the generalization error 
via the R\'enyi mutual information $I_2(X{:}Q)$ between the embedding space 
$Q$ and the classical input space $X$. Thanks to our bound, overfitting 
does not occur when the number of training pairs $T$ is much bigger than $2^{I_2(X{:}Q)}$.
Moreover, we have shown how to bound the approximation error via the mutual information 
between the embedding space and the class space, and shown that the classification error 
can approach it lowest possible value (Bayes risk), in the limit of many measurement shots 
or large Hilbert spaces. 
Our bounds were obtained for the linear loss function, routinely employed in QHT, but different losses can be 
linked to the linear loss via bounds. We have also introduced an information 
bottleneck principle for quantum embeddings, which is independent of the choice of 
loss function and allows us to explore different tradeoffs between accuracy and generalization.

Based on our theoretical results and bounds, we have studied different applications for both 
the classification of quantum and classical data. In particular we have studied the 
classification of the quantum phases of an Ising spin chain and proposed the 
Variational Quantum Information Bottleneck (VQIB) to train quantum embeddings with 
good generalization properties. 

Our analysis can be applied to models of moderate complexity, such as those that can be 
trained with near-term quantum hardware. It is currently an open question 
to understand whether quantum classifiers of very-high complexity can mimic the generalization 
capabilities of classical deep learning. 

\begin{acknowledgements}
L.B.~acknowledges support by the program ``Rita Levi  Montalcini'' for
young researchers. J.P and S.P.~acknowledge  funding from EU Horizon 2020 Research and Innovation Action 
under grant agreement No.~862644 (FET-Open project: ``Quantum Readout Techniques and Technologies'', QUARTET).
L.B.~thanks M.~Schuld and N.~Killoran for helpful comments and discussions about the topics of this paper. 
\end{acknowledgements}

	\appendix

\section{Extended derivation}\label{s:extended}

\subsection{Statistical Learning Theory}

Here we show a brief overview of the tools from statistical learning theory \cite{shalev2014understanding}
that we use throughout the manuscript. As in the previous chapters, we 
assume that there exists an abstract probability distribution that models the inputs and their 
corresponding classes $P(c,x)$. 
This distribution is obviously unknown, but by construction, the samples 
in the training set $\mathcal T$ are drawn independently from $P(c,x)$. 
Suppose now that we have built a classifier $h\in \mathcal H$, where $\mathcal
H$ is the set of classifiers that 
we are considering. We may define the error due to misclassification 
via the loss function $\ell_h(c,x)$, which is zero if and only if $c$ is the correct class of $x$. 
Training is done by minimizing the empirical risk, namely the average loss over the training set 
\begin{equation}
	R^{\mathcal T}(h) = \frac1T\sum_{k=1}^T \ell_h(c_k,x_k)~, 
	\label{e:empirical_risk}
\end{equation}
while the ``true'' risk of a classifier $h$ is given by 
\begin{equation}
	R(h) = \ave_{(c,x)\sim P} [\ell_h(c,x)]~.
	\label{e:risk}
\end{equation}
Supervised learning is practically done via empirical risk minimization, namely the optimal 
data driven classifier is obtained from 
\begin{equation}
	h_\mathcal T = \argmin_{h\in \mathcal H} R^{\mathcal T}(h)~.
\end{equation}
The generalization error 
defines how $h^{\mathcal T}$ performs with unseen data, i.e., 
data not present in the training set. Formally the generalization error is then
defined as $R(h^{\mathcal T})-\inf_{h\in\mathcal H} R(h)$.
Setting $h^*=\argmin_{h\in\mathcal H} R(h)$ as the true optimal classifier, we may bound the generalization error 
$\mathcal G$ as
\begin{align}
	\mathcal G &= R(h^{\mathcal T})-R(h^*) = \\&=\nonumber 
	R(h^{\mathcal T})-R^{\mathcal T}(h^{\mathcal T})+R^{\mathcal T}(h^{\mathcal T})-R^{\mathcal T}(h^*)\\&~~~+
	R^{\mathcal T}(h^*) - R(h^*)  \nonumber
\\&\leq 
	R(h^{\mathcal T})-R^{\mathcal T}(h^{\mathcal T})+ R^{\mathcal T}(h^*) - R(h^*) \nonumber
	\\ &\leq 2\sup_{h\in\mathcal H} \left|R(h)-R^{\mathcal T}(h)\right|, 
	\label{e:maxdev}
\end{align}
where in the first inequality we used the fact that $h^{\mathcal T}$ is optimal for $R^{\mathcal T}$, therefore
$R^{\mathcal T}(h^{\mathcal T})\leq R^{\mathcal T}(h^*) $. The upper bound is known as 
the uniform deviation bound. It represents the maximum deviation between the true and empirical risks, 
Eqs.~\eqref{e:empirical_risk}-\eqref{e:risk}, maximized over the possible classifiers. 

The goal of statistical learning theory is to study how much larger the risk $R(h^{\mathcal T})$ is than 
the Bayes risk, namely $R^{\rm Bayes} = \inf_{h} R(h)$ where the infimum is over all possible hypotheses, 
not restricted to $\mathcal H$. Then by summing and subtracting $R(h^*)$ we get 
\begin{equation}
	R(h^{\mathcal T})-R^{\rm Bayes} = \mathcal G + \mathcal A
\end{equation}
where $\mathcal A= 	R(h^*)-R^{\rm Bayes} $ is the approximation error, which depends on the hypothesis 
space $\mathcal H$. 
One of the central results of statistical learning theory is the following \cite{shalev2014understanding}: %
if $\ell$ has support in $[0,1]$ then with probability at least $1-\delta$ we have that 
\begin{equation}
	\mathcal G 
	\leq 4 \mathcal C_{ T}(\mathcal H) + \sqrt{\frac{2\log(1/\delta)}{T}}~,
	\label{e:genbound}
\end{equation}
where $\mathcal C_{T}(\mathcal H)$ is the Rademacher complexity of $\mathcal H$, which is defined as 
\begin{equation}
	\mathcal C_{T}(\mathcal H) := 
	\ave_{\mathcal T\sim P^T}\ave_{\bs\sigma}
	\left[ \sup_{h\in H}\frac1T\sum_{k=1}^T \sigma_k \ell_h(c_k,x_k) \right],
	\label{e:rademacher}
\end{equation}
where $\sigma_k$ is a random variable which can take two possible values,
$\pm 1$, with the same probability $1/2$,
and the notation $\mathcal T\sim P^T$ means that the $T$ elements in the 
training set $\mathcal T$ are sampled independently from the distribution $P$.
From \eqref{e:genbound} we see that
if the Rademacher complexity of $\mathcal H$ decreases with $T$, then, for sufficiently large $T$, the model is able to generalize 
and correctly predict the class of a new input, not present in the training set $\mathcal T$.

\subsection{Quantum Rademacher Complexity}
Let us calculate the Rademacher complexity of the quantum loss function introduced in 
\eqref{e:qloss}, for which it is clear from the definition that $0\leq	\ell(c_k,x_k) \leq 1$, as requested. For a fixed 
embedding, defining $\mathcal P$ as the set of all possible POVMs, the Rademacher complexity of 
this quantum classifier \eqref{e:qloss} is 
\begin{align}
	\mathcal C_{T}(\mathcal P) &:= \nonumber
	\ave_{\mathcal T\sim P^T}\ave_{\bs\sigma}
	\left[ \sup_{\{\Pi_c\}\in \mathcal P}\frac1T\sum_{k=1}^T \sigma_k \sum_{c\neq c_k}  \Tr[\Pi_c \rho(x_k)]  \right]
													\\ &= 
	\ave_{\mathcal T\sim P^T}\ave_{\bs\sigma}
	\left[ \sup_{\{\Pi_c\}\in \mathcal P}\frac1T\sum_{k=1}^T \sigma_k \Tr[\Pi_{c_k} \rho(x_k)]  \right]
	\label{e:rademacherQ0}
\end{align}
where in the second line we used the second equality in Eq.~\eqref{e:qloss}, the fact that the constant term (from substituting in Eq.~\eqref{e:qloss}) commutes 
with the sup and is averaged out by $\ave_{\bs\sigma}$, and finally the fact that the minus sign 
can be removed by noting that $\bs\sigma$ and $-\bs\sigma$ have the same distribution. 
Let us define 
\begin{equation}
	Q^{\mathcal T}_{c,\bs\sigma} = \frac1T \sum_{k=1}^T \delta_{c,c_k} \sigma_k \rho(x_k)~,
\end{equation}
then by linearity we may rewrite Eq.~\eqref{e:rademacherQ0} as 
\begin{equation}
	\mathcal C_{T}(\mathcal P) := 
	\ave_{\mathcal T\sim P^T}\ave_{\bs\sigma}
	\left[ \sup_{\{\Pi_c\}\in \mathcal P} \sum_c \Tr[\Pi_c Q_{c,\bs\sigma}^{\mathcal T}] \right]~.
	\label{e:rademacherQ}
\end{equation}
In the following sections we show how to bound 	$\mathcal C_{T}(\mathcal P) $ using quantities that 
can be easily computed for a given embedding $x\mapsto\rho(x)$. The main technical result that allows 
such simple expressions is the following 

\begin{lemma}
	Let $A_i$ be a set of operators and $i$  a random variable with probability distribution $p_i$. Then 
	\begin{equation}
		\ave_{i\sim p} \left(\|A_i\|_1\right) \leq \Tr\sqrt{\ave_{i\sim p} \left(A_iA_i^\dagger\right)}~,
			\label{e:ave1norm}
	\end{equation}
where	$\ave_{i\sim p} f(i) := \sum_i p_i f(i)$. 
\end{lemma}
\noindent {\it Proof:} We define the positive operators $X_i:=\sqrt{A_i A_i^\dagger}$. 
Thanks to the definition of trace norm $\|A\|_1= \Tr\sqrt{AA^\dagger}$ and the linearity 
of the trace, it is sufficient to prove that 
\begin{equation}
	\sum_i p_i X_i \leq \sqrt{\sum_i p_i X_i^2}~,
	\label{ee:lemma1eq}
\end{equation}
where the operator inequality $Y\geq 0$ means that $Y$ is a positive operator. 
The above inequality is proven as follows.
Since the function $f(x)=x^2$ is operator convex \cite{carlen2010trace}, we 
may write 
\begin{equation}
	\left(\sum_i p_i X_i\right)^2 \leq \sum_i p_i X_i^2~.
\end{equation}
Moreover, since $g(x)=\sqrt{x}$ is operator monotone \cite{carlen2010trace} we may take
the square root of both sides of the above equation and get \eqref{ee:lemma1eq}. Note that a convex combination of positive matrices is also positive, so the left hand side of \eqref{ee:lemma1eq} is a positive operator, and thus is equal to the square root of its square. This completes 
the proof. $\hfill\square$

We are now ready to write the main result of this section, namely a bound that allows us to express 
the Rademacher complexity of the quantum classifier via the quantity that was introduced in 
\eqref{e:Bdef} and Theorem~\ref{t:main}. We focus on 
binary classification problems, where there are two possible classes, which we call 0 and 1, so 
$c\in\{0,1\}$ and a POVM consists of two positive operators,
$\Pi_0$ and $\Pi_1=1-\Pi_0$. Then, we extend the result to a general multiary classification problem 
with $N_C$ classes.

\begin{thm}\label{t:p2}
	For binary classification problems with fixed embedding $x\mapsto\rho(x)$  and 
	POVM $\mathcal P_{2} = \{\Pi,\openone-\Pi\}$, we find
\begin{equation}
	\mathcal C_T(\mathcal P_2) \leq \frac1{2\sqrt T} \Tr\sqrt{\sum_x P(x)\rho(x)^2} = \frac12\sqrt{\frac{\mathcal B}{T}},
	\label{e:P2bound}
\end{equation}
where $\mathcal B$ was defined in Eq.~\eqref{e:Bdef}. For multiary classification problems with $N$ classes 
we get 
\begin{equation}
	\mathcal C_T(\mathcal P_{N_C}) \leq \sqrt{\frac{N_C \mathcal B}{T}},
	\label{e:PNbound}
\end{equation}
which is slightly larger than \eqref{e:P2bound} when $N_C=2$. 
\end{thm}	

\bigskip

\noindent {\it Proof:} We first focus on binary classification problems. 
Since constant terms are averaged out we can write Eq.~\eqref{e:rademacherQ},
for $\Pi_0=\Pi$ and $\Pi_1=\openone-\Pi$,  as
\begin{equation}
	\mathcal C_T(\mathcal P_2) = 
	\ave_{\mathcal T\sim P^T}\ave_{\bs\sigma} \left[ \sup_{\Pi} 
	\Tr[\Pi(Q_{0,\bs\sigma}^{\mathcal T}-Q_{1,\bs\sigma}^{\mathcal T})] \right]~,
\end{equation}
where again the constant term is averaged out. The maximization over $\Pi$ can be done by adapting 
the Helstrom theorem (see Theorem 13.2 in \cite{bengtsson2017geometry}):
\begin{align*}
	&\ave_{\bs\sigma} \left[ \sup_{\Pi} 
	\Tr[\Pi(Q_{0,\bs\sigma}^{\mathcal T}-Q_{1,\bs\sigma}^{\mathcal T})] \right] \leq \\
	&\frac12\ave_{\bs\sigma}\sup_{\Pi}  \left[ 
	|\Tr[\Pi_0(Q_{0,\bs\sigma}^{\mathcal T}-Q_{1,\bs\sigma}^{\mathcal T})]|  +
|\Tr[\Pi_1(Q_{0,\bs\sigma}^{\mathcal T}-Q_{1,\bs\sigma}^{\mathcal T})]| \right] \leq \\
	&\frac12\ave_{\bs\sigma}\sup_{\Pi}  \left[ 
	\Tr[\Pi_0|Q_{0,\bs\sigma}^{\mathcal T}-Q_{1,\bs\sigma}^{\mathcal T}|]  +
\Tr[\Pi_1|Q_{0,\bs\sigma}^{\mathcal T}-Q_{1,\bs\sigma}^{\mathcal T}|] \right] \leq \\
	&\frac12\ave_{\bs\sigma}
	\Tr[|Q_{0,\bs\sigma}^{\mathcal T}-Q_{1,\bs\sigma}^{\mathcal T}|] ,
\end{align*}
 where $|A|=\sqrt{AA^\dagger}$ and $\|A\|_1=\Tr|A|$. In the first inequality, we are again able to average out the constant term, despite the fact it is within an absolute difference, because setting $\bs\sigma\to-\bs\sigma$ changes its sign whilst not affecting the other term in the absolute difference. The second inequality comes from the fact that $|\Tr[AB]|\leq\Tr[A|B|]$ for any operator B and positive operator A. The third inequality comes from the linearity of the trace and the fact that the elements of the POVM sum to the identity. Therefore,
\begin{align}
	\mathcal C_T(\mathcal P_2) &\leq \ave_{\mathcal T\sim P^T}\ave_{\bs\sigma}
	\left[\frac{\|Q_{0,\bs\sigma}^{\mathcal T}-Q_{1,\bs\sigma}^{\mathcal T}\|_1}2 \right]
	\label{e:C2}
	\\ &= 
	 \ave_{\mathcal T\sim P^T}\ave_{\bs\sigma} 
	 \left\|\frac1{2T}\sum_{k=1}^T(\delta_{c_k,0}-\delta_{c_k,1})\sigma_k \rho(x_k)\right\|_1.
	 \nonumber
\end{align}
An alternative proof of the above inequality without the 1/2, is by using 
the definition \cite{watrous2018theory} of the trace norm 
$\|A\|_1 = \max_{B : \|B\|_\infty \leq 1} \Tr[AB]$, and noting that $\|\Pi\|_\infty\leq 1$ 
for elements of the POVM. 
We now use Eq.~\eqref{e:ave1norm} to get 
\begin{equation}
		\mathcal C_T(\mathcal P_2) \leq \frac12 \Tr\sqrt{
 \ave_{\mathcal T\sim P^T}\ave_{\bs\sigma}
	\left(Q_{0,\bs\sigma}^{\mathcal T}-Q_{1,\bs\sigma}^{\mathcal T}\right)^2}~,
	\label{ee:interm1}
\end{equation}
and by explicit calculation 
\begin{align}
	\left(Q_{0,\bs\sigma}^{\mathcal T}-Q_{1,\bs\sigma}^{\mathcal T}\right)^2 &= 
	\frac1{T^2}	
\sum_{k,j=1}^T(\delta_{c_k,0}-\delta_{c_k,1})(\delta_{c_j,0}-\delta_{c_j,1})\times 
\nonumber\\ &\quad\quad 
\sigma_k\sigma_j \rho(x_k) \rho(x_j).
\end{align}
Since the $\sigma_j$s are independent and with zero mean, we get 
\begin{align}
	 &\ave_{\mathcal T\sim P^T}\ave_{\bs\sigma}\left(Q_{0,\bs\sigma}^{\mathcal T}-Q_{1,\bs\sigma}^{\mathcal T}\right)^2 = 
	 \ave_{\mathcal T\sim P^T}	\sum_{k=1}^T\frac{(\delta_{c_k,0}-\delta_{c_k,1})^2}{T^2} \rho(x_k)^2 
	 \nonumber 
 \\& =\quad\quad\quad\ave_{(c,x)\sim P}\left[ \frac{1}{T} \rho(x)^2\right]~,
 \label{e:aveTsigma}
\end{align}
where we used the fact that $(\delta_{c_k,0}-\delta_{c_k,1})^2=1$ and that
$(c_k,x_k)$ are independent and identically distributed. 
Inserting the above equation into \eqref{ee:interm1} and using $P(x)=\sum_c P(c,x)$ we get 
\eqref{e:P2bound}, which completes the first part of the theorem.

For the multiary classification problem with $N_C$ equiprobable classes, 
using H\"older's inequality in \eqref{e:rademacherQ}, and noting that
$\|\Pi_c\|_\infty\leq 1$, we may write
\begin{align}
	\mathcal C_{T}(\mathcal P_{N_C}) &\leq 
	\ave_{\mathcal T\sim P^T}\ave_{\bs\sigma}
	\sum_c \|Q_{c,\bs\sigma}^{\mathcal T}\|_1
														 \\&\leq
	N_C \ave_{\mathcal T\sim P^T}\ave_{\bs\sigma}\ave_c
	\|Q_{c,\bs\sigma}^{\mathcal T}\|_1
														 \\&\leq
	N_C \Tr \sqrt{ 
	\ave_{\mathcal T\sim P^T}\ave_{\bs\sigma}\ave_c
(Q_{c,\bs\sigma}^{\mathcal T})^2}
														 \\&\leq
N_C \Tr \sqrt{ \frac1{N_C T} \sum_{c,x} P(c,x) \rho(x)^2} 
														 \\&\leq \sqrt{\frac{N_C}T}
\Tr \sqrt{ \sum_{x} P(x) \rho(x)^2},
\end{align}
where in the second line we have substituted the sum over $c$ with an average where 
$c$ is sampled from the uniform distribution with $p_c=1/N_C$, in the third line 
we use Eq.~\eqref{e:ave1norm}, in the fourth line we perform the averages as 
in Eq.~\eqref{e:aveTsigma}, and in the last line we simply employ the marginal distribution, 
as in Eq.~\eqref{e:P2bound}.  
$\hfill\square$

\subsection{Bound on the approximation error}

In this section 
we focus on the approximation error \eqref{e:Adef} and 
prove the following important result. 

\begin{thm}\label{t:approx}
	Given some $x$-dependent states $\rho(x)$, if  we define an embedding using $N$ copies 
	$x\mapsto \rho(x)^{\otimes N}$, then $\mathcal A\to 0$ for $N\to \infty$ as long as 
	$F(\rho(x),\rho(y))\neq 0$ for $x\neq y$. Moreover, 
	\begin{equation}
		\lim_{N\to\infty} \frac{\log \mathcal A}{N}  \leq \log F_{\rm max}
		\label{e:asint}
	\end{equation}
	where $F_{\rm max} = \max_{x\neq y} F(\rho(x),\rho(y))$. 
\end{thm}

\noindent
{\it Proof:} 
From the definition of the approximation error $\mathcal A=R(h^*)-R^{\rm Bayes}$, we write
\begin{align}
	R&= \sum_x \sum_{c\neq c'} P(c,x) \Tr[\Pi^*_{c'} \rho(x)],
	\\
	R^{\rm Bayes}&= \sum_x \sum_{c\neq c'} P(c,x) \ell^{\rm Bayes}(c',x),
\end{align}
where $\ell^{\rm Bayes}(c,x) = \delta_{c,b(x)}$ and $b(x)=\argmax_c P(c|x)$. Note that the second summation is over all $c$ and $c'$ such that $c\neq c'$.
Therefore,
\begin{align}
	\mathcal A  = \sum_x \sum_{c\neq c'} P(x,c)\left(\Tr [\Pi^*_{c'}\rho(x)]-\delta_{c',b(x)}\right)
\end{align}
The approximation error is calculated using the optimal measurement for a given encoding $\rho(x)$, however we can upper bound it by replacing this optimal measurement with a suboptimal measurement. 
We may find a suboptimal strategy as follows: we know that there always exists some POVM $\Pi_x$ that obeys \cite{banchi2020quantum,montanaro2019pretty}
\begin{equation}
	q(x) \Tr[\Pi_y \rho(x)] \leq \sqrt{q(x) q(y)} F(\rho(x),\rho(y)),
	\label{e:pgm}
\end{equation}
for any probability distribution $q(x)$. From $\Pi_x$ we can then construct the POVM $\Pi_c$ as follows 
\begin{equation}
	\Pi_c = \sum_x \delta_{c,b(x)}\Pi_x,
\end{equation}
namely we first try to learn the value of $x$, and then perform the standard Bayesian classification 
to get the class. 
Now there are many possibilities to find bounds depending on the choice of $q(x)$ in \eqref{e:pgm}. 
Here we chose to use $q(x)=p(x|c)$, namely we train the measurements to recognise all inputs within 
a certain class, and then check whether the Bayes classifier predicts a different result.
We may write 
\begin{align}
	\mathcal A  &\leq \sum_x \sum_{c\neq c'} P(c)P(x|c)\left[\sum_y \delta_{c',b(y)}\Tr[\Pi_y\rho(x)] 
	-\delta_{c',b(x)}\right]
		\nonumber \\&=
		\sum_x \sum_{c\neq c'} P(c)\left[\sum_y \delta_{c',b(y)}\mathcal F_{xy}^c
-P(x|c)\delta_{c',b(x)}\right] \nonumber 
						 \\ &=
						 \sum_{x\neq y} \sum_{c\neq b(y)} P(c) \mathcal F_{xy}^c,
						 \label{e:Abound}
\end{align}
where, in the last line, the summation is over all $x$ and $y$ such that $x\neq y$, and where
\begin{equation}
	F_{xy}^c = \sqrt{P(x|c)P(y|c)}F(\rho(x),\rho(y)).
\end{equation}
The upper bound \eqref{e:Abound} is typically too large to be practical. 
However, it can be used to show an important result. If we define an embedding
as $x\mapsto \rho(x)^{\otimes N}$, then $F_{xy}^c = \sqrt{P(x|c)P(y|c)}F(\rho(x),\rho(y))^N \to 0$ 
for $N\to \infty$ as long as $F(\rho(x),\rho(y))\neq 0$. Moreover, since $F(\rho(x),\rho(y))\leq F_{\rm max}$, 
we get Eq.~\eqref{e:asint}.
$\hfill\square$

\bigskip

Thanks to the above theorem, we see that taking 
copies of a simple embedding guarantees that $\mathcal A\to 0$ for $N\to\infty$, as we 
observe in the numerical simulations shown in Fig.~\ref{fig:angle}.

\section{Further inequalities}\label{s:further}
In this section we discuss other inequalities and connections with other entropic quantities. We first 
recall the following inequality, which will be extensively used in this section:
\begin{equation}
	\Tr\sqrt{\sum_i X_i} \leq \sum_i \Tr \sqrt{X_i},
	\label{e:sqrtineq}
\end{equation}
which is valid for any set of positive operators $X_i$ 
 (see Ref.~\cite{audenaert2014upper} 
for a proof).

We first discuss the risk \eqref{e:qloss} and empirical risk \eqref{e:emplossPi} for multiary classification 
problems with $N_C$ classes. In this case there is no known analytic form of the optimal POVM, but suboptimal 
choices can be constructed using pretty good measurements 
 \cite{barnum2002reversing,banchi2020quantum}: calling $T_c$ the number of samples in the training set 
 with class $c$, we may write 
$R^{\mathcal T}(\Pi) = 1-\sum_c \frac{T_c}T \Tr[\Pi_c \rho_c^{\mathcal T}]$ and the error for an 
optimal measurement can be bounded as \cite{banchi2020quantum,barnum2002reversing} 
\begin{equation}
	R^{\mathcal T}(\Pi^{\mathcal T}) \leq \sum_{c\neq c'}\frac{\sqrt{T_c
	T_{c'}}}T F(\rho_c^{\mathcal T},\rho_{c'}^{\mathcal T}),
	\label{e:trainerrmulti}
\end{equation}
where $F(\rho,\sigma)=\|\sqrt{\rho}\sqrt{\sigma}\|_1$ is the quantum fidelity. Using the strong 
concavity of the fidelity 
\begin{equation}
	\mathbb{E}^{\mathcal T} R^{\mathcal T}(\Pi^{\mathcal T}) \leq \sum_{c\neq c'}\frac{\sqrt{T_c
	T_{c'}}}T F(\rho_c,\rho_{c'}),
\end{equation}
which is a multiclass generalization of \eqref{e:averisk}. Therefore, even for classification 
problems with $N_C>2$ low risk is possible when the fidelity between the average states with 
inputs belonging to different classes is low.

As for the generalization error, we see that the complexity 
$\mathcal B$ defined in Eq.~\eqref{e:Bdef} does not depend on the classes. Nonetheless, using 
\eqref{e:sqrtineq} we also get the following inequalities 
\begin{align}
	\mathcal B &\leq \left(\sum_c \sqrt{P(c) \mathcal B_c}\right)^2,\\
	\mathcal B_c&= \left(\Tr\sqrt{\sum_x P(x|c) \rho(x)^2}\right)^2.
	\label{e:Bc}
\end{align}
Setting $\sigma_c = \sum_x P(x|c) \rho(x)^2 = \sum_{i=1}^{r_c} 
\lambda_i \ket{\lambda_i}\!\bra{\lambda_i}$, where $r_c$ is the rank of $\sigma_c$, 
we find 
\begin{align}
	\mathcal B_c = \left(\sum_i \sqrt{\lambda_i}\right)^2 = \sum_i \lambda_i + 
	\sum_{i\neq j} \sqrt{\lambda_i\lambda_j}.
	\label{e:asdassd}
\end{align}
Thanks to Jensen's inequality, for every set of positive $x_i$ of size $n$, we can write
$\sum_{i=1}^n\sqrt{x_i} \leq \sqrt{ n \sum_{i=1}^n x_i}$. 
In Eq.~\eqref{e:asdassd}, the number of terms in the second sum is $r_c^2-r_c$, so 
\begin{align}
	\mathcal B_c &\leq
	\sum_i\lambda_i +\sqrt{(r_c^2-r_c)
	\sum_{i\neq j} \lambda_i\lambda_j}
						\\ &= \sum_i\lambda_i +
						\sqrt{(r_c^2-r_c)\left(\sum_{i,j} \lambda_i\lambda_j -\sum_i\lambda_i^2\right)}
						\\ &= \Tr[\sigma_c] + 
						\sqrt{(r_c^2-r_c)\left(\Tr[\sigma_c]^2 -\Tr[\sigma_c^2]\right)}.
\end{align}
For pure state embeddings, $\sigma_c=\rho_c$ and we 
get  \eqref{e:boundpurity}.

Another interesting bound can be found by applying the Cauchy-Schwarz inequality
$\Tr[\sqrt{X}]^2\leq \Tr[X]\Tr[\openone]$ to \eqref{e:Bdef}. We find
\begin{equation}
	\mathcal B \leq D \sum_x P(x) \Tr[\rho(x)^2] \leq D,
\end{equation}
where $D$ is the dimension of the embedding Hilbert space.

We now study the embedding $\rho(x)$ using tools 
from quantum information. We define an extended tripartite mixed $\rho_{CXQ}$ state as in Eq.~\eqref{e:rhoextended},
where $CX$ is the data Hilbert space, $C$ is spanned by the classes $\ket c$ and  $X$ by the labels $\ket x$, 
while $Q$ is the Hilbert space of the quantum embedding $\rho(x)$. 
We now introduce the R\'enyi conditional mutual information of $\rho_{CXQ}$ following Prop.~8 
from \cite{berta2015renyi}
\begin{align}
	\label{e:renyicond}
	&I_{ \alpha}(X{:}Q|C) = \\ &~ \frac\alpha{\alpha-1} \log_2\Tr\left(\left[\rho_{C}^{\frac{\alpha-1}2}
	\Tr_X \left(\rho_{CX}^{\frac{1-\alpha}2} \rho_{CXQ}^\alpha
\rho_{CX}^{\frac{1-\alpha}2} \right)\rho_{C}^{\frac{\alpha-1}2}\right]^{\frac1\alpha}\right),
\nonumber
\end{align}
where  $\rho_{CX} = \Tr_Q[\rho_{CXQ}]=\sum_{xc}P(c,x)\ket{cx}\!\bra{cx}$, and
$\rho_C=\Tr_X[\rho_{XC}] = \sum_c P(c) \ket c\!\bra c$. When multiplying operators that span different Hilbert spaces, it is implicit that the operators take a tensor product with the identity on the spaces that they do not span (e.g., $\rho_{CX}\rho_{CXQ}=(\rho_{CX}\otimes\openone_Q)\rho_{CXQ}$).
By explicit computation 
\begin{align}
	&\Tr_X \left(\rho_{CX}^{\frac{1-\alpha}2} \rho_{CXQ}^\alpha\rho_{CX}^{\frac{1-\alpha}2} \right) = 
\\ &~~~~~ \nonumber 
	\sum_{cx} P(c,x)^{1-\alpha} P(c,x)^\alpha \ket c\!\bra c \otimes \rho(x)^\alpha=
\\ &~~~~~ \nonumber 
	\sum_{cx} P(c,x) \ket c\!\bra c \otimes \rho(x)^\alpha,
	\\ 
	 &\rho_{C}^{\frac{\alpha-1}2}
	\Tr_X \left(\rho_{CX}^{\frac{1-\alpha}2} \rho_{CXQ}^\alpha
\rho_{CX}^{\frac{1-\alpha}2} \right)\rho_{C}^{\frac{\alpha-1}2}= 
\\ &~~~~~ \nonumber 
\sum_{cx} P(c,x) P(c)^{\alpha-1}\ket c\!\bra c \otimes \rho(x)^\alpha=
\\ &~~~~~ \nonumber 
\sum_{cx} P(x|c) P(c)^{\alpha}\ket c\!\bra c \otimes \rho(x)^\alpha
\end{align}
and 
\begin{align}
	\label{e:condinfo}
	I_{\alpha}&(X{:}Q|C) =\\ &\nonumber   \frac\alpha{\alpha-1} 
	\log_2\left[\sum_c P(c) \Tr \left( \Big[\sum_x P(x|c) \rho(x)^\alpha\Big]^{\frac1\alpha}\right)\right].
\end{align}
We note a similarity between the above expression and the quantities that are found in the generalization 
bound \eqref{e:Bc}. Indeed, for a uniform distribution over $N_C$ classes, $P(c) = 1/N_C$, i.e., when 
all classes are equally likely, we find from \eqref{e:Bc} that 
\begin{equation}
	\mathcal B \leq {2^{I_{2}(X{:}Q|C)}N_C},
	\label{e:Bcondinfo}
\end{equation}
and thus show a direct link between the generalization bound and the
R\'enyi conditional mutual information of $\rho_{CXQ}$. 
Therefore, good generalization is possible whenever $I_{2}(X{:}Q|C)$ is small or for large training sets $T\gg 2^{I_{2}(X{:}Q|C)}N_C$. 

We can interpret the space $Q$ in \eqref{e:rhoextended} as compression of the input into 
a quantum state. Assuming that the conclusions from Ref.~\cite{fischer2020conditional} (which were originally formulated for the classical Shannon entropy) can be trivially extended 
to R\'enyi entropies, optimal compression happens when 
\begin{equation}
	I_{\alpha}(X{:}Q|C) = 
	I_{\alpha}(X{:}C|Q) = 
	I_{\alpha}(C{:}Q|X) = 0.
\end{equation}
A zero conditional mutual information means that the three systems form a Markov chain: 
 conditioning over one of the three systems makes the other two mutually independent. 
Quantum mechanically $	I(X{:}Q|C) =0$ if $\rho_{XQC}=\rho_{XC}^{1/2}\rho_C^{-1/2}\rho_{QC}\rho_C^{-1/2}\rho_{XC}^{1/2}$ 
\cite{berta2015renyi}, which is not generally satisfied by the state \eqref{e:rhoextended}. 
According to Ref.~\cite{fischer2020conditional}, we should both minimize 	$I(X{:}Q|C) $ and maximize 
$I(C{:}Q) $.
The R\'enyi generalization of $I(C{:}Q)$ can be obtained using a similar expression to \eqref{e:renyicond}, by applying the expression for conditional mutual information to a state that is independent of the conditioning system (i.e., by calculating $	I(C{:}Q|X)$ for the state $\rho_{CQ}\otimes\rho_X$). We get the expression
\begin{align}
	I_{\alpha}(C{:}Q) &= \frac\alpha{\alpha-1} \log_2\Tr\left(\left[
	\Tr_C \left(\rho_{C}^{\frac{1-\alpha}2} \rho_{CQ}^\alpha
\rho_{C}^{\frac{1-\alpha}2} \right)\right]^{\frac1\alpha}\right)
		\nonumber \\ & =
\frac\alpha{\alpha-1} \log_2\Tr\left(\Big[
\sum_c P(c) \rho_c^\alpha\Big]^{\frac1\alpha}\right)~,
\end{align}
where $\rho_c$ was introduced in Eq.~\eqref{e:rhoc}. Simpler expressions that are 
directly connected with the fidelity may be obtained when for $\alpha=1/2$.
For pure state embeddings $\rho(x)^\alpha=\rho(x)$ and we get 
\begin{align}
	I_{1/2}(X{:}Q|C) &= -\log_2 \sum_c P(c) \Tr\left(\Big[\sum_x P(x|c)\sqrt{\rho(x)}\Big]^2\right)
	\nonumber \\ &= -\log_2 \sum_c P(c) \mathcal F(c),
\end{align}
where 
\begin{equation}
\mathcal 	F(c) =	\sum_{x,y} P(x|c)P(y|c) F(\rho(x),\rho(y))^2,
\end{equation}
is the average squared fidelity between embeddings of inputs from the same class
$c$. Therefore,  $\mathcal F(c)$ should be 
maximized to minimize $	I_{1/2}(X{:}Q|C) $. In other words, low conditional mutual information is possible 
when $F(\rho(x),\rho(y))\simeq 1$ for $x$ and $y$ belonging to the same class. 
Moreover, 
\begin{align}
	-I_{1/2}(C{:}Q) &= \log_2 \Tr\left[\sum_c P(c) \sqrt{\rho_c}\right]^2
						\\  &= \log_2 \left[\sum_c P(c)^2 + \sum_{c\neq c'} P(c)P(c')\mathcal F(c,c')\right],
\nonumber
\end{align}
where 
\begin{equation}
	\mathcal F(c,c') = \Tr[\sqrt{\rho_c}\sqrt{\rho_{c'}}]\leq F(\rho_c,\rho_{c'}) 
\end{equation}
is related to the fidelity between the average states $\rho_c$ and $\rho_{c'} $ for different 
classes. Since the logarithm is monotonic, the minimization of 	$-I_{1/2}(C{:}Q) $ is possible 
by minimizing $F(\rho_c,\rho_{c'})$. Therefore, using R\'enyi entropies with $\alpha=1/2$, 
we recover the same conclusions as in the previous section:
\emph{ a good embedding is one for which the fidelity 
between two embedded states is small if the inputs are from different classes and high if the inputs 
are from the same class. }

\section{Connection to kernels}\label{s:kernels}

We prove the following theorem that allows us to express functions on ensemble of pure states 
\begin{lemma}\label{l:f}
	If the function $f:[0,1]\mapsto\mathbb R$ admits a series representation, then 
	\begin{equation}
		\Tr f\left(\sum_x p(x) \ket{\psi(x)}\!\bra{\psi(x)}\right) = \Tr f(\tilde K),
	\end{equation}
	where the operator $\tilde K$ has entries 
	\begin{equation}
		\tilde K_{x,y} = \sqrt{p(x)p(y)} \bra{\psi(x)}\psi(y)\rangle.
		\label{e:appker}
	\end{equation}
\end{lemma}
\noindent The proof trivially follows from the  matrix power
\begin{align}
	&\Tr \left(\sum_x p(x) \ket{\psi(x)}\!\bra{\psi(x)}\right)^n =\\ &=  \sum_{ \{x_j\} } \prod_{j=1}^n 
	p(x_j) \bra{\psi(x_n)}\psi(x_1)\rangle 
	\cdots \bra{\psi(x_{n-1})}\psi(x_n)\rangle 
\nonumber 	\\ & = \nonumber 
	\sum_{ \{x_j\} } \tilde K_{x_n,x_1} \tilde K_{x_1,x_2}\cdots \tilde K_{x_{n-1},x_n} = \Tr \tilde K^n.
\quad\quad\quad\square
\end{align}

When  $|z-1|<1$, 
the square root function admits the expansion $\sqrt z = \sum_{k=0}^\infty \frac{(-1)^k}{k!}\left(-\frac12\right)_k
(z-1)^k$, where $(a)_k = a(a+1)\cdots(a+k-1)$ is the Pochammer symbol. Since  $(z-1)^k$ can be 
expressed as a sum of $z^n$ for $n\leq k$, we can apply the above lemma to show that for ensemble of pure states 
\begin{equation}
	\mathcal B = \left(\Tr \sqrt{\tilde K}\right)^2 =  \left(\Tr \sqrt K\right)^2,
\end{equation}
which is Eq.~\eqref{e:kernel}. 
Note that the matrices $\tilde K$ and $K$ are related by a similarity transformation and,
accordingly, have the same eigenvalues. 

For almost diagonal kernel matrices, 
we may write $\tilde K=K_d+K_o$ where $(K_d)_{xy}=p(x)\delta_{xy}$ is the diagonal part and $K_o$ the off-diagonal part,
as in Eq.~\eqref{e:appker} with $x\neq y$. If the off-diagonal elements are much smaller $(\mathcal O(\epsilon))$ 
than the diagonal ones, then we may expand $\sqrt{\tilde K} = \sqrt{K_d} + \epsilon K_1
+ \epsilon^2 K_2 +\mathcal O(\epsilon^3)$. Taking the square 
on either sides we can find 
\begin{align}
	(K_1)_{xy} &= \frac{(K_o)_{xy}}{\sqrt{p(x)} + \sqrt{p(y)}}, \\ 
	(K_2)_{xy} &=- \frac{(K^2_1)_{xy}}{\sqrt{p(x)} + \sqrt{p(y)}},
\end{align}
and 
\begin{equation}
	\Tr\sqrt K = 2^{H_2(X)/2} - \frac14 \sum_{x\neq y} \frac{\sqrt{p(x)p(y)}}{\sqrt{p(x)}+\sqrt{p(y)}} F_{xy }^2
 + \mathcal O(\epsilon^4),
\end{equation}
where $F_{xy}= |\bra{\psi(x)}\psi(y)\rangle|$ and $2^{H_2(X)} = \Tr[\sqrt{K_d}]$.

\end{document}